\documentclass[12pt]{article}

\usepackage{amsmath,amssymb}

\title{Valuations on Functionally Closed Sets of 
Quantum Mechanical Observables 
and Von Neumann's `No-Hidden-Variables' Theorem\footnote{Forthcoming 
in Dennis Dieks and Pieter Vermaas (eds.), \emph{The 
Modal Interpretation of Quantum Mechanics}, University of Western 
Ontario Series in Philosophy of Science, Kluwer Academic 
Publishers, 1998.}}
\author{Jason Zimba and Rob Clifton}
\date{}
\begin{document}
\raggedbottom
\maketitle

\section{Introduction}

Every modal interpretation of quantum mechanics has the following distinctive
feature:

\begin{quote}

Given the (pure or mixed) quantum state $W$ of a system with Hilbert space
$\sf H$, the interpretation specifies those self-adjoint operators on $\sf H$
which correspond to observables with {\bf definite values} in state $W$.

\end{quote}

\noindent We are not asserting that all realist interpretations of quantum mechanics
must necessarily do this, nor are we asserting that doing this in itself
counts as giving an interpretation.  But certainly the central task of modal
interpretations is to provide an ontology of the properties of quantum
systems that circumvents the measurement problem, without falling prey to the
various `no-hidden-variables' theorems.  So, to accomplish that task, modal
interpretations must tell us which observables of a system we can and should
be realists about.  Morever, this must at least involve specifying which of a
system's discrete-valued observables can be said to possess definite values
statistically distributed in conformity with the statistics prescribed by the
density operator $W$ of the system.  Our main aim in this paper is to take a
detailed look at some of the mathematical issues that arise naturally in the
context of such a specification.

For \emph{continuous-valued} observables, the notion of `possessing a definite
value' may need to be replaced by something like `possessing a value lying in
(or restricted to) a definite interval.'  Furthermore, although our
mathematical analysis will indeed apply when ${\sf H}$ is
infinite-dimensional, a few of our results remain sensitive to the difference
between discrete- and continuous-spectrum observables on ${\sf H}$.  Thus our
analysis (both conceptual and mathematical) will be complete only with
respect to the notion of `possessing a definite value' appropriate to
observables with a discrete spectrum.  Of course, since modal interpretations
have so far only been rigorously developed for such observables, this will
not hamper the application of our results to them.  But there is clearly more
work to be done (for recent progress in this connection, see Clifton [1997]).

If at a certain instant of time the state of a system is $W$, then we shall
denote the set of definite-valued observables of the system by ${\bf D}(W)$,
or simply ${\bf D}$.  In purely \emph{mathematical} discussions of ${\bf D}$,
we shall take as given that its observables are represented by self-adjoint
operators, and we shall refer to ${\bf D}$ as the system's set of
definite-valued \emph{operators}.

For our purposes it will prove useful to ask the following question: \emph{a
priori}, what sort of mathematical structure, if any, is it natural to
attribute to $\bf D$?  Is $\bf D$ a (real) vector space, in which case real
linear combinations of definite-valued operators are necessarily
definite-valued?  An algebra of some kind, in which case polynomials
involving definite-valued operators are definite-valued?\  Does it matter if
the operators in question commute?  And finally, is it helpful to view
``functional closure'' properties like these as normative requirements on
possible modal interpretations?

In section 2 we shall define a few of these functional closure properties
more carefully, drawing attention to some the mathematical issues that come
into play when we prescribe them for $\bf D$.  Such functional closure issues
figure prominently, for example, in von Neumann's [1955]
`no-hidden-variables' theorem -- where it is assumed that any real linear
combination of operators in $\bf D$ will itself be in $\bf D$, regardless of
whether these operators are compatible.

Apart from making this `structural' requirement, von Neumann's theorem also
contains an assumption about the \emph{values} possessed by the observables in
$\bf D$; specifically, it assumes that these possessed values must obey the
same polynomial relationships as do the corresponding observables -- again,
regardless of whether these observables are compatible.  The received view,
first clearly articulated by Bell [1966], is that the acceptability of the
theorem as a `no-go' result is undercut at this point by the lack of
attention von Neumann paid to compatibility.  Thus Bell argued that in the
case of \emph{incompatible observables}, it is \emph{not} reasonable to require
of any hidden-variable theory that its value assignments necessarily reflect
the observables' algebraic relationships.

The received view, then, is that von Neumann's functional requirement for
possessed values is so strong that the theorem fails to rule out hidden
variables in any convincing way.  However, our own diagnosis of what makes
the theorem unacceptable will be somewhat different.  In fact, in most of
what follows, we shall take the bold step of adopting functional requirements
that are (in a sense) even \emph{stronger} than von Neumann's polynomial ones.

In the first place, we shall require that \emph{any} self-adjoint function of
observables in $\bf D$ must itself be in $\bf D$ -- again,
\emph{irrespective} of the compatibility of the observables.  Having adopted
this requirement, the latter part of section 2 will be devoted to isolating a
simple necessary and sufficient condition on the \emph{projection} operators
in $\bf D$ for $\bf D$ to be functionally closed in this strong sense.
Interestingly, the projection sets specified as definite-valued by a number
of proposed modal interpretations all meet this condition; hence we are able
to show that according to all of those interpretations, arbitrary functions
of definite-valued operators are themselves definite-valued.

Then in section 3 we turn to the issue of the \emph{values} of the observables
in $\bf D$.  This is where von Neumann's no-go theorem packs its punch.  If,
for example, one assumes that $\bf D$ is the set of \emph{all} self-adjoint
operators on $\sf H$, then it is easy to show, as von Neumann did, that no
assignment of values to the observables in $\bf D$ can respect their
polynomial functional relations.  But modal interpretations are not so
liberal about what they take $\bf D$ to be.  Because they take their sets of
definite-valued observables to be a certain kind of \emph{sub}algebra of the
set of all self-adjoint operators on $\sf H$, we shall show that there do
indeed exist valuations on their definite-valued sets $\bf D$ which respect
polynomial relationships among the observables in $\bf D$.  Moreover -- and
here is where we make the second of our strengthened functional requirements
-- we shall show that even if we require that the valuations respect 
\emph{arbitrary} functional relationships among the observables in $\bf D$ (again
regardless of whether the observables commute), then there are still enough
of them to represent the statistics prescribed by quantum mechanics for
observables in $\bf D$, as measures over the available `functional'
valuations.  Thus we locate the fault in von Neumann's theorem, not directly
in his assumption that valuations must always respect this or that type of
functional relationship, but rather in his \emph{tacit} assumption that 
\emph{every} self-adjoint operator may be considered a candidate for an
element of $\bf D$.

Section 3 ends with the primary mathematical result of the paper:  a simple
condition on the projections in a functionally closed set $\bf D$ which is
necessary and sufficient for $\bf D$ to support enough functional valuations
to represent quantum statistics.

In sections 2 and 3, which form the main part of the paper, a number of
mathematical concepts will need to be invoked.  Section 2 draws on the theory
of von Neumann algebras, and section 3 draws on the lattice-theoretic idea of
a quasiBoolean algebra (first introduced in Bell and Clifton [1995]).  But
our exposition will be self-contained, all of the mathematics needed (most of
it well-known) will be introduced \emph{en route}, and the theorems we prove
will be understandable by anyone who has followed our mathematical
definitions and terminology (most of it standard).

In section 4 we bring things to a close by amplifying the above remarks on
the relevance of our results to von Neumann's theorem.  One point to be made
in this respect is that since modal interpretations can recover quantum
statistics, they provide an existence proof that all the explicitly stated
demands placed by von Neumann on `hidden-variable theories' can be met (save
his tacit, and by no means compelling, assumption that every observable has a
value).  And having thereby circumvented von Neumann's theorem, modal
interpretations also automatically circumvent all `no-go' theorems that
attempt to strengthen the case against `hidden variables' by making weaker
assumptions than von Neumann did -- most notably the theorems of Jauch and
Piron [1963] and Kochen and Specker [1967].

\section{Functional Closure Properties for Sets of Definite-valued
Observables}

\subsection{Degrees of Functional Closure}

Here are four properties that interpreters might consider attributing to the
set of definite-valued self-adjoint operators $\bf D$ on a Hilbert space $\sf
H$.  (Note that we shall always assume that $\bf D$ contains the identity
operator.)

\begin{itemize}

\item \emph{Compatible polynomial $^{\ast}$-closure}.  We will say that $\bf
D$ has \emph{compatible polynomial $^{\ast}$-closure} if, whenever the
commuting operators $Q$ and $S$ are in $\bf D$, the operators $a Q + S$ and
$QS$ are also in $\bf D$, for all real $a$.  (To put it another way, $\bf D$
has compatible polynomial $^{\ast}$-closure if any real polynomial function
of commuting operators in $\bf D$ is also in $\bf D$.  In this case one might
call $\bf D$ a \emph{partial real algebra}.)
\item \emph{Compatible $^{\ast}$-closure}.  We will say that $\bf D$ has
\emph{compatible $^{\ast}$-closure} if, whenever the commuting operators
$\{Q_{\alpha}\}$ are in $\bf D$, any self-adjoint operator that is a (not
necessarily polynomial) function of the $Q_{\alpha}$ is in $\bf D$.  (For
finite-dimensional $\sf H$, this is equivalent to compatible polynomial
$^{\ast}$-closure. Note that a function is self-adjoint if it maps a set of self-adjoint operators to a self-adjoint operator.)
\item \emph{Polynomial $^{\ast}$-closure}.  We will say that $\bf D$ has 
\emph{polynomial $^{\ast}$-closure} if any self-adjoint polynomial function of
operators in $\bf D$ is also in $\bf D$.  (In this case one might call $\bf
D$ a \emph{real algebra}.)
\item $^{\ast}$-\emph{Closure}.  We will say that $\bf D$ has 
$^{\ast}$-\emph{closure} if, whenever the operators $\{Q_{\alpha}\}$ are in
$\bf D$, any self-adjoint operator that is a (not necessarily polynomial)
function of the $Q_{\alpha}$ is in $\bf D$.
\end{itemize}

A brief word on the \emph{star} in `$^{\ast}$-closure'.  Generally speaking,
we are considering what it means for a set of operators to be closed under
functional operations.  When we come to spelling out how an arbitrary (not
necessarily polynomial or self-adjoint) function of a set of operators is defined (i.e.\ in the
next subsection), it will turn out that the question of the functional
closure of a set of operators has everything to do with the question of
whether the set is \emph{topologically} closed, in an appropriate topology.
We will need to have a different notation for these two closure concepts in
order to discuss their relationship.

There are grounds to think that in any reasonable interpretation, the set of
definite-valued observables ought, at least, to have compatible polynomial
$^{\ast}$-closure.  The \emph{orthodox} (Dirac-von Neumann) interpretation,
for example, is certainly one in which the set of definite-valued observables
has this property.  This is because to an orthodox interpreter, if
$\{Q_{\alpha}\}$ is a set of definite-valued operators, then the state vector
must be an eigenvector of each $Q_{\alpha}$ in the set.  But in that case,
the state vector will clearly also be an eigenvector of any polynomial
function of the $Q_{\alpha}$.  Hence according to the orthodox
interpretation, any self-adjoint polynomial function of definite-valued
operators is itself definite-valued.

To refuse to attribute compatible polynomial $^{\ast}$-closure to the set of
definite valued operators, one would have to believe something like the
following:  that in some situations a particle could, for example, have a
definite value of energy without having a definite value of energy-squared.
One way to believe this would be to deny that operators like
``energy-squared'' represent physical quantities in the first place, though
it is not clear what extra insights on the problem that would bring.  But in
any case, it would seem that in order to dispute the \emph{a priori}
reasonableness of compatible polynomial $^{\ast}$-closure, one would have to
adopt what is in some ways an extremely conservative viewpoint.

On the other end of the spectrum, an extremely liberal interpreter might be
unsatisfied with a condition as \emph{weak} as compatible, polynomial
$^{\ast}$-closure.  Such an interpreter might even be willing to entertain
the idea that in any reasonable interpretation, the set of definite-valued
operators should be nothing less than $^{\ast}$-closed (e.g., see Clifton
[1995a,b]).  Perhaps this goes too far.  But for those who are tempted to
consider $^{\ast}$-closure to be an \emph{outlandish} requirement, we shall be
showing that a large group of modal interpretations do in fact satisfy it,
along with the orthodox interpretation and, of course, the naive realist
interpretation (`every observable has a definite value').

We shall henceforth be adopting $^{\ast}$-closure as a requirement on $\bf
D$, partly because $^{\ast}$-closure is compatible with so many proposed
interpretations, and partly because the requirement of $^{\ast}$-closure
places a number of useful mathematical tools at our disposal.  Using these
tools, we shall translate the condition of $^{\ast}$-closure on $\bf D$ into
a simple equivalent condition on the set of \emph{projections} in $\bf D$.
This condition will doubtless prove useful for generating new modal
interpretations that, by construction, are manifestly functionally closed.
(For a further discussion of the issues raised by various
functional closure requirements, see Zimba [1998].)

As outlined in the introduction, another reason for focusing on
$^{\ast}$-closed sets of definite-valued observables is that, by leading us
to a class of modal interpretations that easily circumvent von Neumann's
`no-hidden-variables' theorem, they allow us to stress that the difficulty
with this theorem does not have to be seen as stemming solely from concerns
about the functionality of valuations for incompatible observables.

\subsection{Von Neumann Algebras and $^{\ast}$-Closure}

We begin by summarizing some elementary notions concerning `functions of
operators' which will elucidate the concept of $^{\ast}$-closure.  We
consider only bounded linear operators on the Hilbert space $\sf H$.

\begin{itemize}
\item \emph{Strong limit of a sequence of operators}. Consider a sequence
$\{G_{n}\}$ of operators.  Suppose that for each vector $x$ there exists a
vector $y_{x}$ such that 
\[
\lim_{n\rightarrow \infty}\| G_{n} x-y_{x}\| = 0.
\]
Then the map $x\stackrel{G}{\mapsto} y_{x}$ is said to be the \emph{strong
limit} of the sequence $\{G_{n}\}$:
\[
\lim_{n\rightarrow \infty}G_{n} = G.
\]
It follows that if the $G$ defined above exists, then it is unique, and
linear if the $G_{n}$ are.  (These facts are easy to prove using the triangle
inequality.)

(There are two other common notions of the limit of a sequence of operators:
a stronger notion, called the \emph{uniform} limit, and a weaker notion,
called the \emph{weak} limit.  We shall not be explicitly considering either
of these, though in all the cases we are concerned with the weak and strong
limits coincide. For a fuller discussion of some of the conceptual issues at stake here, see Clifton [1997] and Zimba [1998].)

\item \emph{Polynomial function}. A \emph{polynomial function} of the
operators in $\{Q_{\alpha}\}$ is a finite linear combination of products of
powers of the $Q_{\alpha}$, with complex coefficients.

\item \emph{Operator-valued function of operators}.  An operator $G$ is said
to be a \emph{function} of the operators in $\{Q_{\alpha}\}$ if it is the
strong limit of a sequence of polynomial functions of the $Q_{\alpha}$.
(This recalls the approach of ordinary analysis, in which functions are often
defined as infinite series -- or, in other words, as limits of sequences of
polynomials.)

\end{itemize}

In the hope that it will make the mathematics easier to read, we shall use
the following font conventions:

\begin{itemize}

\item Calligraphic capital:  A set of operators.  For example,
${\cal B}$.

\item Bold-face capital:  A set of specifically \emph{self-adjoint} operators.
For example, $\bf D$.

\item Capital:  An operator.  For example, $Q$.

\item Lower-case italics:  A complex scalar or vector, depending on context.
For example, $a$ or $x$.

\end{itemize}

More definitions:

\begin{itemize}

\item \emph{Self-adjoint set}. If a set of operators ${\cal B}$ contains
$Q^{\dagger}$ whenever it contains $Q$, then it is called a 
\emph{self-adjoint} set.  (We use $Q^{\dagger}$ for the adjoint instead of
$Q^{\ast}$ to avoid confusing a `$^{\ast}$-closed set' with a `self-adjoint
set.'  Note also the distinction between the phrases ``a set of self-adjoint
operators'' and ``a self-adjoint set of operators''!)

\item \emph{$^{\ast}$-algebra}.  A self-adjoint set ${\cal B}$ is called a
$^{\ast}$-algebra if it contains $aQ + T$ and $QT$, where $a$ is any complex
scalar, whenever it contains $Q$ and $T$.  (In other words, a self-adjoint
set is a $^{\ast}$-algebra if it contains all polynomial functions of its
members.)

\item \emph{von Neumann algebra}. A $^{\ast}$-algebra ${\cal A}$ is called a
\emph{von Neumann algebra} if it contains the identity and is closed in the
strong operator topology -- that is, if strongly convergent sequences of
operators in ${\cal A}$ converge to operators in ${\cal A}$.  To put it
another way, a $^{\ast}$-algebra ${\cal A}$ is a von Neumann algebra if it
contains the identity and if any function of operators in ${\cal A}$ is also
in ${\cal A}$.  (We have required that ${\cal A}$ contain the identity in
order to simplify our presentation, but this requirement is not part of the
standard definition.)
\item \emph{Commutant}.  The \emph{commutant} of a set of operators ${\cal
B}$ is the set of all operators on $\sf H$ that commute with all operators in
${\cal B}$.  We use a prime to denote the commutant:
\[
{\cal B}^{\prime}= \{T:TB=BT\;\;\mbox{\rm for all}\;\; B\in {\cal B}\}.
\]
It follows that ${\cal A}\subseteq{\cal B}$ implies ${\cal B}^{\prime}
\subseteq {\cal A}^{\prime}$ and that $({\cal A}\cup {\cal B})^{\prime} =
{\cal A}^{\prime}\cap{\cal B}^{\prime}$. Furthermore, ${\cal B}^{\prime}$
will be a $^{\ast}$-algebra whenever ${\cal B}$ is a self-adjoint set.

We write the \emph{second} commutant $({\cal B}^{\prime})^{\prime}$ as ${\cal
B}^{\prime\prime}$  (So: an operator $Q$ is in ${\cal B}^{\prime\prime}$ if
it commutes with any operator that commutes with every operator in ${\cal
B}$.) It is then elementary to show that ${\cal B}\subseteq{\cal
B}^{\prime\prime}$ and ${\cal B}^{\prime} = {\cal B}^{\prime\prime\prime}$
for any operator set ${\cal B}$.
\end{itemize}

This last notion of the commutant of a set of operators is especially useful
for elucidating $^{\ast}$-closure.  Given a set of operators ${\cal B}$, ask
yourself what kinds of operators ${\cal B}^{\prime\prime}$ contains (apart
from those in ${\cal B}$ itself).  Well, suppose an operator $T$ commutes
with everything in ${\cal B}$.  Then $T$ certainly commutes with any
polynomial function of operators in ${\cal B}$.  So any polynomial function
of operators in ${\cal B}$ commutes with any operator $T$ that commutes with
every operator in ${\cal B}$.  In other words, any polynomial function of
operators in ${\cal B}$ is contained in ${\cal B}^{\prime\prime}$. (Note that
these polynomial functions need not be self-adjoint.)  Hence ${\cal
B}^{\prime\prime}$ is an algebra.

What's more, if ${\cal B}$ is a \emph{self-adjoint} set, then ${\cal
B}^{\prime\prime}$ will also be a self-adjoint set. This follows as a result
of the fact that self-adjointness of sets is preserved under the operation of
taking the commutant.  For suppose that ${\cal B}$ is a self-adjoint set, and
consider any $T$ in ${\cal B}^{\prime}$.  Then for any $B$ in ${\cal B}$, we
have $B^{\dagger}\in{\cal B}$, so $[T,B^{\dagger}]=0$.  Taking adjoints, we
have $[T^{\dagger},B]=0$.  Since $B$ was arbitrary, we conclude that
$T^{\dagger}\in{\cal B}^{\prime}$.  And since $T$ was arbitrary, we conclude
that ${\cal B}^{\prime}$ is self-adjoint.  Hence ${\cal B}^{\prime}$ is
self-adjoint whenever ${\cal B}$ is, which was to be shown.  Together with
the fact that ${\cal B}^{\prime\prime}$ is always an algebra, we see that if
${\cal B}$ is a \emph{self-adjoint} set, then ${\cal B}^{\prime\prime}$ will
be a $^{\ast}$-\emph{algebra}.

Summarizing then, for a self-adjoint set ${\cal B}$, the set ${\cal
B}^{\prime\prime}$ is a $^{\ast}$-algebra generated by ${\cal B}$, containing,
for example, all polynomial functions of operators in ${\cal B}$.  What's
more, the following remarkable theorem of von Neumann shows that ${\cal
B}^{\prime\prime}$ contains \emph{all} functions of operators in ${\cal B}$:
\begin{itemize}
\item \emph{The Double Commutant Theorem} (von Neumann).  

Let ${\cal A}$ be a
\mbox{$^{\ast}$-algebra}.  Then ${\cal A}$ is a von Neumann algebra (closed in the
strong operator topology and containing the identity) if and only if ${\cal
A} = {\cal A}^{\prime\prime}$.  (Topping [1971])
\end{itemize}
Since ${\cal B}^{\prime} = {\cal B}^{\prime\prime\prime}$, we have ${\cal
B}^{\prime \prime} = ({\cal B}^{\prime\prime})^{\prime\prime}$ from which it
follows that for a self-adjoint set ${\cal B}$, the set ${\cal
B}^{\prime\prime}$ is a von Neumann algebra.  In fact, ${\cal
B}^{\prime\prime}$ is the smallest von Neumann algebra containing ${\cal B}$.
To see this, suppose that ${\cal A}$ is a von Neumann algebra containing
${\cal B}$; so ${\cal B}\subseteq{\cal A}$.  Then ${\cal
A}^{\prime}\subseteq{\cal B}^{\prime}$, whence ${\cal
B}^{\prime\prime}\subseteq{\cal A}^{\prime\prime}$.  But since ${\cal A}$ is
a von Neumann algebra, we have ${\cal A} = {\cal A}^{\prime\prime}$, and we
may therefore conclude ${\cal B}^{\prime\prime}\subseteq{\cal A}$.  Thus any
von Neumann algebra containing ${\cal B}$ also contains ${\cal
B}^{\prime\prime}$, so that ${\cal B}^{\prime\prime}$ is the smallest von
Neumann algebra containing ${\cal B}$ -- which is to say, ${\cal
B}^{\prime\prime}$ is the smallest $^{\ast}$-algebra containing ${\cal B}$
and strong limits of sequences of operators in ${\cal B}$.  Hence:
\begin{quote}
\emph{The von Neumann algebra} ${\cal B}^{\prime\prime}$ 
\emph{generated by a self-adjoint set} ${\cal B}$ \emph{is the set of all
functions of operators in} ${\cal B}$.
\end{quote}

Now let's return to $^{\ast}$-closure.  We said that a set of self-adjoint
operators $\bf D$ is $^{\ast}$-\emph{closed} if it contains all \emph{self-adjoint}
functions of operators in $\bf D$.  So the important difference between
$^{\ast}$-closure and \emph{topological} closure in the strong operator
topology is that $^{\ast}$-closure refers only to self-adjoint functions of
operators.  (Hence the star.)  To relate the two notions precisely, we make
the following definition:
\begin{itemize}
\item \emph{Self-adjoint part.}  The \emph{self-adjoint part} of a
$^{\ast}$-algebra ${\cal A}$ is the set ${\bf S}({\cal A}) = \{ Q\in{\cal A}:
Q = Q^{\dagger} \}$.
\end{itemize} 
With this, we can relate $^{\ast}$-closure to topological closure in the
strong operator topology in a way that should now be obvious:
\\[\baselineskip]
{\bf Theorem 1}.  Let $\bf D$ be a set of self-adjoint
operators.  Then $\bf D$ is
$^{\ast}$-closed
if and only if it is the self-adjoint part of a von Neumann
algebra ${\cal A}$.
(Symbolically, ${\bf D} = {\bf S}({\cal A})$.)
\\[\baselineskip]
{\bf Proof}. $(\Rightarrow)$  This direction says that if $\bf D$ is
$^{\ast}$-closed, then $\bf D$ is the self-adjoint part of a von Neumann
algebra.  To prove this, first recall what has just been said: that the von
Neumann algebra ${\bf D}^{\prime\prime}$ is the set of all functions of
operators in $\bf D$.  In particular then, the \emph{self-adjoint} functions
of operators in $\bf D$ are just the operators in $\bf S({\bf
D}^{\prime\prime})$.  If these are assumed to be contained in $\bf D$, then
we must have $\bf D\supseteq{\bf S}({\bf D}^{\prime\prime})$. And it is
obvious that for a set of self-adjoint operators, ${\bf S}({\bf
D}^{\prime\prime})\supseteq\bf D$.  So we see that if $\bf D$ is
$^{\ast}$-closed, then $\bf D = {\bf S}({\bf D}^{\prime\prime})$.  In other
words, a $^{\ast}$-closed set of self-adjoint operators is the self-adjoint
part of the von Neumann algebra it generates.

($\Leftarrow$)  This direction says that if ${\cal A}$ is a von Neumann
algebra, then its self-adjoint part ${\bf S} ({\cal A})$ is $^{\ast}$-closed.
To prove this, we need to show that strongly convergent sequences of
self-adjoint operators in ${\cal A}$ converge to self-adjoint operators.  To
this end, suppose ${\cal A}$ is a von Neumann algebra, and consider the set
${\bf D} = {\bf S}({\cal A})$.  If $\{Q_{n}\}$ is a strongly convergent
sequence of operators in $\bf D$, then $\{Q_{n}\}$ is also a strongly
convergent sequence of operators in the closed set ${\cal A}$; hence
$\{Q_{n}\}$ converges strongly to some $Q\in{\cal A}$.  We need to show that
the limit operator $Q$ is self-adjoint, so that it lies in $\bf D$.

For any vector $x$, define a sequence of real numbers $\{q_{n}(x)\}$ by
$q_{n}(x) = \langle x, Q_{n}x\rangle$. Also, define $q(x)$ to $\langle x,
Qx\rangle$.  Then, making use of the Schwarz Inequality, we have
\begin{eqnarray*}
|q_{n}(x) - q(x)| & = & |\langle x, Q_{n}x\rangle - \langle x, Qx\rangle |\\
& = & |\langle x, (Q_{n} - Q)x \rangle |\\ & \leq & \parallel x \parallel
\cdot \parallel (Q_{n}-Q)x\parallel\\ & \rightarrow & 0
\end{eqnarray*}
by strong convergence of the sequence $\{Q_{n}\}$ to $Q$.  This shows that
$q_{n}(x)\rightarrow q(x)$.  But it is elementary to show that if a sequence
of complex numbers $q_{n}$ converges in modulus to a complex number $q$, then
the real and imaginary parts of $q_{n}$ converge separately to the real and
imaginary parts of $q$.  Since each $q_{n}$ is real, this means that the
limit of our sequence, $q(x) \equiv \langle x,Qx\rangle$, must also be real.
Hence the limit operator $Q$ has real expectation values on every vector $x$,
from which it follows that $Q$ is self-adjoint.  Thus $\bf D$ is
$^{\ast}$-closed.\ \emph{QED}.

\subsection{Projection Operators and $^{\ast}$-Closure}

With Thm.~1 we have a criterion for deciding when a definite-valued set $\bf
D$ is $^{\ast}$-closed.  But in most of the literature, modal interpretations
are defined from the perspective of \emph{idempotent} observables, i.e.\
projections.  In this approach
\begin{enumerate}
\item We specify a set $\bf d$ of projections with definite values in the
state $W$;
\item  We adopt the condition that a self-adjoint operator is in ${\bf D}$ if
and only if all the spectral projections of the operator are in ${\bf d}$.
\end{enumerate}
For now, we want to allow $\bf D$ to contain observables with continuous
spectra.  So if a self-adjoint operator has a continuous spectrum, we shall
extend standard terminology and take the `spectral projections' of the
operator to be the set of all projections of the form $P - Q$, where $PQ = Q$
and both $P$ and $Q$ are in the spectral family of the operator.  (Thus the
`spectral projections' are the projections associated with the various 
\emph{ranges} of values the observable can take up.)

Rule 2 is often tacit in the literature, but it is usually what is intended.
In fact, 2 follows from requiring $^{\ast}$-closure of $\bf D$.  For, by the
spectral theorem, a self-adjoint operator can be approximated as closely as
one likes by an appropriate real linear combination of its spectral
projections, and conversely, each such projection is a characteristic
function of the operator.

The procedure of specifying $\bf D$ by specifying the subset $\bf d$ of its
projections and adopting rule 2 is at the heart of what one might call ``the
projection operator approach to the problem of definiteness.'' Using this
approach, a number of modal interpretations, along with the naive realist and
orthodox interpretations, can be characterized as follows.  Let the density
matrix for the system be $W$, with spectral resolution ${\bf X} =
\{X_{i}\}$.  So $X_{i}X_{j} = \delta_{ij}X_{j}$, $\sum X_{i} = I$, and $\sum
w_{i}X_{i} = W$. (And let $X_{0}$ denote the projection onto the
null space of $W$, if it has a non-trivial null space.) In the special case
where $W$ is a pure state represented by a unit vector $\psi$, let $\{
P_{\psi_{R_{j}}} \}$ be the projection operators associated with the
one-dimensional subspaces generated by the (non-zero) components of $\psi$
that lie in the eigenspaces $\{R_{j}\}$ of an observable $R$ (with discrete
spectrum).  In this notation, the definite-valued projections of a number of different modal interpretations are given by
\begin{eqnarray*}
{\bf d}_{NR}& = &\{P^{2} = P = P^{\dagger} \} \\
{\bf d}_{B} & = & \{P^{2} = P = P^{\dagger}: P P_{\psi_{R_{j}}} =
P_{\psi_{R_{j}}} \; \mbox{or} \; 0 \; \mbox{for all} \; j \}\\
{\bf d}_{C} & = & \{P^{2} = P = P^{\dagger}: PX_{i} =
X_{i} \; \mbox{or} \; 0\; \mbox{for all} \; i \neq 0 \}\\
{\bf d}_{K,D}& =& \{P^{2} = P = P^{\dagger}: PX_{i} =
X_{i} \; \mbox{or} \; 0 \;
\mbox{for all} \; i \} \\
{\bf d}_{O}& = &\{P^{2} = P = P^{\dagger}: PW = W \;
\mbox{or} \; 0 \}.
\end{eqnarray*}
Roughly speaking, we have ordered these projection sets from `largest' to
`smallest'.  At the top of the list is the naive realist, who considers 
\emph{every} projection to have a definite value.  The three proposals in the
middle, due to Bub [1997], Clifton [1995a], and Kochen [1985] and Dieks
[1995], are more discriminating.  They consider a projection $P$ to be
definite-valued whenever it ``resolves'' the projections in a certain
orthogonal set into two classes: those whose ranges are contained in that of
$P$, and those whose ranges are orthogonal to that of $P$.  (${\bf d}_{C}$ is
closely related to ${\bf d}_{K,D}$ and, in fact, is called the `Kochen-Dieks'
interpretation by Clifton [1995a].  The difference is that since ${\bf
d}_{K,D}$ includes $X_{0}$ in its definition, it must form a Boolean algebra
of projections -- the Boolean algebra generated by the $\{X_{i}\}$, which sum
to the identity operator.)
Most parsimonious is the orthodox interpreter, who does not permit the
projections in the spectral resolution of $W$ to be ``resolved'' in this way.
According to the orthodox view, in order for a projection to have a definite
value, the projection must either annihilate $W$ or preserve it.

In each of these interpretations, the set of projections is expressed in
terms of a smaller set ${\bf X}$.  We generalize this notion as follows:
\begin{itemize}
\item $\bf X$-\emph{form set}.  We shall say that a set of projections $\bf d$
is an $\bf X$-\emph{form} set if there is a \emph{mutually orthogonal} set of
projections $\bf X$, not containing the zero projection, in terms of which
$\bf d$ can be written as
\[
{\bf d} = \{P^{2} = P = P^{\dagger}: PX = X \; \mbox{or} \; 0 \; \mbox{for
all} \; X \in {\bf X} \}.
\]
Equivalently, one may say that $\bf d$ is an $\bf X$-\emph{form} set if there
is a subset ${\bf X}\subseteq {\bf d}$, not containing the zero projection,
in terms of which ${\bf d}$ can be expressed as above.
\end{itemize}
Except for ${\bf d}_{NR}$, all the projection sets above are $\bf X$-form
sets (noting that ${\bf X}_{O} = \{\sum_{i\neq 0}X_{i}\})$.  $\bf X$-form
sets are what Dickson [1995a,b] calls Faux-Boolean algebras, and he shows
that they have desirable properties in addition to those we shall stress
here.  Our focus will be on the fact that an $\bf X$-form set generates a
$^{\ast}$-closed set of definite-valued observables, and that an $\bf X$-form
set guarantees the existence of sufficiently many valuations on that
$^{\ast}$-closed set to justify the name `definite-valued.'

In order to talk about the sets of definite-valued operators ${\bf D}_{NR}$,
${\bf D}_{B}$, ${\bf D}_{C}$, etc. corresponding to ${\bf d}_{NR}$, ${\bf
d}_{B}$, ${\bf d}_{C}$, etc., we make two natural definitions:
\begin{itemize}
\item \emph{Restriction}.  Given a set of self-adjoint operators ${\bf D}$,
define the \emph{restriction} of $\bf D$ to be the set of idempotent members
of $\bf D$.  We denote the restriction by $\underline{\bf D}$.  (We shall
also use the notation $\underline{\cal B}$  for the set of all projections in
an arbitrary set of operators $\cal B$.)
\item \emph{Extension}.  Given a set of projections $\bf d$, define the
\emph{extension} of $\bf d$ as follows.  A self-adjoint operator is in the
extension if and only if all its spectral projections lie in $\bf d$.
Denote the extension of $\bf d$ by $\overline{\bf d}$.
\end{itemize}
Note that the extension is not defined to include only \emph{discrete}
observables with spectral projectors in $\bf d$ (recall our earlier
generalization of the terminology `spectral projections' to cover the
continuous case).  When we need to confine ourselves to sets $\bf d$ with
extensions containing only discrete observables (and two of our main results
below are, so far, limited to that case), we shall say so explicitly.

Let $\bf d$ be a set of projections, with $\overline{\bf d}$ its extension.
With Thm.~1 we have a test of whether $\overline{{\bf d}}$ is
$^{\ast}$-closed.  We now convert that into a test given directly in terms of
projections and $\bf d$ itself: \\[\baselineskip] 
{\bf Theorem 2}.  Given a set of projections $\bf d$, its extension
$\overline{\bf d}$  is $^{\ast}$-closed if and only if $\bf d$ is the
restriction of the commutant of some set of projections $\bf P$.
(Symbolically, $\overline{\bf d}$ is $^{\ast}$-closed iff ${\bf d} =
\underline{{\bf P}^{\prime}}$.) \\[\baselineskip]
{\bf Proof}.  We saw in Thm.~1 that $\overline{\bf d}$ is $^{\ast}$-closed if
and only if $\overline{\bf d}= {\bf S}({\cal A})$ for some von Neumann
algebra ${\cal A}$.  We first show that $\overline{\bf d}= {\bf S}({\cal A})$
is equivalent to ${\bf d}= \underline{\bf S}({\cal A})$. It is easy to see
that $\underline{({\overline{\bf d}})}= {\bf d}$, so it suffices to show
$\overline{{\underline{\bf S}}({\cal A})} = {\bf S}({\cal A})$.

Let $Q$ be an operator in $\overline{{\underline{\bf S}}({\cal A})}$.  Then
by definition the spectral projections of $Q$ are contained in
$\underline{\bf S}({\cal A})$, and hence in ${\cal A}$.  Now, $Q$ may be
approximated as closely as desired by an appropriate linear combination of
these spectral projections; in other words, $Q$ is the strong limit of a
sequence of operators in ${\cal A}$.  But since ${\cal A}$ is a von Neumann
algebra, ${\cal A}$ \emph{contains} its (strong) limits; this means that $Q$
itself must be in ${\cal A}$.  Moreover, since $Q$ is self-adjoint, $Q$ must
actually be in ${\bf S}({\cal A})$.  This shows that
$\overline{{\underline{\bf S}}({\cal A})} \subseteq {\bf S}({\cal A})$.

Conversely, let $Q$ be an operator in ${\bf S}({\cal A})$.  Then of course
$Q$ is in ${\cal A}$.  Therefore, since each projection in the spectral
family of $Q$ is a characteristic function of $Q$, and since ${\cal A}$ is a
von Neumann algebra (hence functionally closed), each spectral projection of
$Q$ must also be in ${\cal A}$.  Clearly then each spectral projection must
be in $\underline{\bf S}({\cal A})$.  Thus all of the spectral projections of
$Q$ are in $\underline{\bf S}({\cal A})$, which is to say that $Q$ is in
$\overline{{\underline{\bf S}}({\cal A})}$.  This shows that ${\bf S}({\cal
A})\subseteq\overline{{\underline{\bf S}}({\cal A})}$, so\, that\,\,
$\overline{{\underline{\bf S}}({\cal A})} = {\bf S}({\cal A})$
as claimed.

So $\overline{\bf d}$ is $^{\ast}$-closed if and only if there is a von
Neumann algebra ${\cal A}$ for which ${\bf d} = {\bf\underline{S}}({\cal A})
= \underline{\cal A}.$ We now show that there exists such a von Neumann
algebra if and only if there is a set of projections $\bf P$ for which ${\bf
d} = \underline{{\bf P}^{\prime}}$.

If $\bf P$ is a set of projections then it is self-adjoint, in which case
${\bf P}^{\prime}$ is a $^{\ast}$-algebra containing the identity.  And since
${\bf P}^{\prime} = {\bf P}^{\prime\prime\prime}$, ${\bf P}^{\prime}$ is
therefore a von Neumann algebra (by the Double Commutant Theorem).  This
establishes that if there is a set of projections ${\bf P}$ for which ${\bf
d}= \underline{{\bf P}^{\prime}}$, then $\bf d$ is the restriction of a von
Neumann algebra.

Conversely, if there is a von Neumann algebra ${\cal A}$ for which ${\bf d} =
\underline{\cal A}$, then define ${\bf P} = \underline{{\cal A}^{\prime}}$.
We show that whenever ${\cal A}$ is a von Neumann algebra
$\underline{(\underline{{\cal A}^{\prime}})^{\prime}} =
\underline{\cal A}$ so that $\underline{{\bf P}^{\prime}} = {\bf d}$.

Let $T$ be any operator (not necessarily self-adjoint) in ${\cal
A}^{\prime\prime}$, where ${\cal A}$ is \emph{any} set of operators.  Then $T$
commutes with everything in ${\cal A}^{\prime}$, so of course $T$ commutes
with everything in $\underline{{\cal A}^{\prime}}$.  By definition then T is
in $(\underline{{\cal A}^{\prime}})^{\prime}$ , and we have established (for
\emph{any} set of operators ${\cal A}$) that ${\cal A}^{\prime\prime}\subseteq
(\underline{{\cal A}^{\prime}})^{\prime}$.

Next, let $T$ be any operator (not necessarily self-adjoint) in
$(\underline{{\cal A}^{\prime}})^{\prime}$, where ${\cal A}$ now is any 
\emph{self-adjoint} set of operators.  This means that $T$ commutes with all
the projections in ${\cal A}^{\prime}$.  Consider then an arbitrary
self-adjoint operator $Q$ in ${\cal A}^{\prime}$.  Since ${\cal A}$ is a
self-adjoint set, ${\cal A}^{\prime}$ is a von Neumann algebra, so all of the
spectral projections of $Q$ must be contained in ${\cal A}^{\prime}$.  Since
$T$ must commute with each of these spectral projections, $T$ must therefore
commute with $Q$ itself.  In other words, from the fact that $T$ is in
$(\underline{{\cal A}^{\prime}})^{\prime}$ we may conclude that $T$ commutes
with everything in ${\bf S} ( {\cal A}^{\prime})$.

Next, since ${\cal A}^{\prime}$ is a $^{\ast}$-algebra, any operator
$V\in{\cal A}^{\prime}$ can be written as $V=V_{R} + i V_{I}$, where $V_{R} =
(V + V^{\dagger})/2\in {\bf S}({\cal A}^{\prime})$ and $V_{I} = -i (V -
V^{\dagger})/2 \in {\bf S}({\cal A}^{\prime})$.  So if $T$ commutes with
everything in ${\bf S}({\cal A}^{\prime})$, then in fact $T$ commutes with
everything in ${\cal A}^{\prime}$.  Thus $T$ is in ${\cal A}^{\prime\prime}$,
so that $(\underline{{\cal A}^{\prime}})^{\prime}\subseteq {\cal
A}^{\prime\prime}$, and we have finally shown (for any self-adjoint set of
operators ${\cal A}$) that $(\underline{{\cal A}^{\prime}})^{\prime} = {\cal
A}^{\prime\prime}$.
Consequently, for any \emph{von Neumann algebra} ${\cal A}$, we will clearly
have $(\underline{{\cal A}^{\prime}})^{\prime} = {\cal A}$.  And this of
course implies $\underline{(\underline{{\cal A}^{\prime}})^{\prime}} =
\underline{\cal A}$.  So if $\bf d$ is of the form ${\bf d} = \underline{\cal
A}$ for some von Neumann algebra ${\cal A}$, then there is a set of projections $\bf
P$ (namely $\underline{{\cal A}^{\prime}}$) for which ${\bf d} =\underline{{\bf
P}^{\prime}}$. \ \emph{QED}.\\[\baselineskip]\indent
With this theorem we can quickly show that under modal interpretations, as
well as under the orthodox interpretation, arbitrary self-adjoint functions
of definite-valued operators are themselves definite-valued.  (This is
trivially true for the naive realist interpretation.)
\\[\baselineskip]
{\bf Corollary}.  If $\bf d$ is of $\bf X$-form, then $\overline{{\bf d}}$ is
$^{\ast}$-closed.\\[\baselineskip]
{\bf Proof}.  Suppose $\bf d$ is of $\bf X$-form for some set $\bf X$, and
define
\[
{\bf P} \equiv \{P^{2} = P = P^{\dagger}:  XP =P \; \mbox{for some} \; X \in 
{\bf X}\}.\]
We show that $\bf d$ coincides with $\underline{{\bf P}^{\prime}}$, hence by
the previous theorem $\overline{\bf d}$ is \mbox{$^{\ast}$-closed}.

Consider any projections $P$ in $\bf P$ and $Q$ in ${\bf d}$; so for some $X$
in $\bf X$, $XP = P$ and $QX = X$ or $0$.  It follows that $QP =P$ or $0$, so
that $Q$ commutes with $P$.  Therefore every projection in $\bf d$ commutes
with every element of $\bf P$, and we have ${\bf d}\subseteq \underline{{\bf
P}^{\prime}}$.

Conversely, suppose a projection $Q$ commutes with every $P$ in $\bf P$,
i.e.\ suppose $Q$ is in $\underline{{\bf P}^{\prime}}$.  Since ${\bf X}$ is a
subset of $\bf P$, $Q$ commutes with every $X$.  To conclude $\underline{{\bf
P}^{\prime}} \subseteq {\bf d}$   we must show more, namely that $QX = X$ or
$0$ (for any $X$).

Therefore consider the operators $QX$ and $(I-Q)X$.  Since $[Q,X] = 0$, these
are orthogonal projections that sum to $X$.  If they are \emph{both}
non-zero, then there are normalized, orthogonal vectors $v$ and $w$ with $v$
in the range of $QX$ and $w$ in the range of $(I-Q)X$.  Write this as $v \in
\mbox{ran}(QX)$ and $w\in \mbox{ran}((I-Q)X)$. Now consider the vector $z = v
+ w$ and its associated one-dimensional projection $Z$.  Clearly $z \in
\mbox{ran}(X)$, so $XZ =Z$; consequently $Z\in  {\bf P}$.  But note also that
$[Z, Q]\neq 0$, since $z$ is not in $\mbox{ran}(Q)$ or $\mbox{ran}(I-Q)$.
This contradicts the initial assumption that $Q$ is in $\underline{{\bf
P}^{\prime}}$.  Hence at least one of $QX$ or $(I-Q)X$ must be $0$.
\ \emph{QED}.\\[\baselineskip]

Thm.~2 allows us to say something more specific about the structure of a set
$\bf d$ of projections with $^{\ast}$-closed extension, viz.\ about its
lattice-theoretic structure.  We first recall the relevant aspects of lattice
theory.
\begin{itemize}
\item \emph{Lattice}. A \emph{lattice} is a partially ordered set $L$ in
which each pair of
elements $x, y\in L$ has a supremum or \emph{join} -- denoted by 
$x \vee y$ -- and an infimum or \emph{meet} -- denoted by $x\wedge y$.  (We shall be
dealing only with lattices which
have a maximum element 1, and a minimum element $0$.)
\item \emph{Completeness}.  A lattice $L$ is \emph{complete} if every subset
of $L$ has both a join and a meet in $L$.
\item \emph{Ortholattice}.  A lattice $L$ is orthocomplemented, or an
\emph{ortholattice}, if every $x\in  L$ has a complement $x^{\perp}\in L$ 
satisfying:
\begin{eqnarray*}
x \vee x^{\perp} & = &1\\
x \wedge x^{\perp} & = &0\\
x \leq y & \Rightarrow & y^{\perp}\leq x^{\perp}\\
(x^{\perp})^{\perp} & = & x.
\end{eqnarray*}
\item \emph{Orthomodular lattice}.  An ortholattice is \emph{orthomodular}
if in addition it
satisfies:
\[
x \leq y \Rightarrow  y = x \vee (y \wedge x^{\perp}).
\]
\item \emph{Atom}.  An \emph{atom} of a lattice $L$ is a minimal non-zero
element.  That is, $x$ is an atom of $L$ if $x\neq 0$ and if, for all $y \in
L$, $y \leq x$ implies $y = x$ or $y =0$.
\item \emph{Atomic}.  A lattice $L$ is \emph{atomic} if for all non-zero
$y\in L$ there is an atom $x\in L$ where $x\leq y$.
\end{itemize}
Note that if $L$ is a complete, atomic, orthomodular lattice, then every
element of $L$ is the join of the atoms contained in that element, i.e.\ for
any \mbox{$y\in L$,} $y = \vee A$ where $A = \{ x \in L: x \leq y$ and $x$ is an
atom$\}$.  The proof is straightforward:  for any $y\in  L$, clearly $\vee A
\leq y$ (noting $\vee A\in L$, by completeness).  So by orthomodularity, $y =
(\vee A)\vee [y \wedge (\vee A)^{\perp}]$.  But $y\wedge (\vee A)^{\perp} =
0$, otherwise, by atomicity of $L$, the set $A$ would not exhaust the atoms
contained in $y$.  (In the proof of Thm.~4, later on, we shall invoke this
result without comment.)

The set of all projections on a Hilbert space $\sf H$ forms a lattice $L({\sf
H})$.  Since the projections $P$ are in one-to-one correspondence with the
closed subspaces $\mbox{ran}(P)$ onto which they project, the projections may
be ordered by ordering their ranges by inclusion.

Given two projections $P$ and $Q$ on $\sf H$, their meet $P\wedge Q$ is
defined to be the projection onto $\mbox{ran}(P)\cap \mbox{ran}(Q)$ -- a
well-defined notion since the intersection of closed subspaces is itself a
closed subspace.  And their join $P\vee Q$ is defined to be the projection
onto the norm-closed span of $\mbox{ran}(P)\cup \mbox{ran}(Q)$.  For
\emph{arbitrary} sets of projections $\{P_{\alpha}\}$, the existence of
$\wedge\{P_{\alpha}\}$ follows from the fact that for any set of closed
subspaces $\{\Pi_{\alpha}\}$, there is a largest closed subspace $\Pi$
contained in each $\Pi_{\alpha}$ (Topping [1971]).  Then
$\wedge\{P_{\alpha}\}$ is defined to be the projection onto $\Pi$.  Similar
remarks hold for $\vee\{P_{\alpha}\}$, and $L({\sf H})$ is thus a
\emph{complete} lattice.

Identifying 1 with the identity operator and $0$ with the zero operator, and
associating to each projection $P$ the projection $P^{\perp}$ onto the
orthocomplement of $\mbox{ran}(P)$ -- which is a closed subspace -- $L({\sf
H})$ becomes a complete orthomodular lattice.  $L({\sf H})$ is also atomic,
with its atoms being the projections onto the one-dimensional subspaces of
${\sf H}$.

As we show next, much the same is true for any \emph{subset} of projections
on $\sf H$, provided the subset has a $^{\ast}$-closed extension -- such a
subset always picks out a complete, orthomodular \emph{sublattice} of $L({\sf
H})$.  (Atomicity will be discussed shortly.)\\[\baselineskip] 
{\bf Theorem 3}.  Given a subset of projections ${\bf d} \subseteq L({\sf
H})$, its extension $\overline{{\bf d}}$ is $^{\ast}$-closed only if ${\bf
d}$ forms a complete, orthomodular sublattice of $L({\sf
H})$.\\[\baselineskip] 
{\bf Proof}.  If $\overline{{\bf d}}$ is $^{\ast}$-closed, then Thm.~2 says
that ${\bf d}$ is given by $\underline{{\bf P}^{\prime}}$, where $\bf P$ is
some set of projections on $\sf H$.  So we must show that $\underline{{\bf
P}^{\prime}}$ forms a complete orthomodular lattice.

Let $\bf Q$ be any subset of projections in $\underline{{\bf P}^{\prime}}$
and let $\wedge {\bf Q}$ be the meet (in $L({\sf H})$) of all the elements in
$\bf Q$.

\underline{Claim}: If an operator $A$ commutes with every projection in $\bf
Q$, i.e.\ if $A \in {\bf Q}^{\prime}$, then $A$ commutes with 
$\wedge {\bf Q}$.

To see this, let $r$ be a vector in $\mbox{ran}(\wedge {\bf Q})$.  Then for any $P
\in {\bf Q}$ we have
\[
PAr = APr
\]
(by assumption)
\[
\Rightarrow \qquad P(Ar) = (Ar)
\]
($r \in {\rm ran}(\wedge {\bf Q}) \subseteq {\rm ran} (P)$)
\[
\Rightarrow \qquad Ar \in {\rm ran}(P).
\]
But since this holds for all $P \in {\bf Q}$, we must therefore also have
\[
Ar \in {\rm ran}(\wedge {\bf Q}).  
\]
Thus whenever $r \in {\rm
ran}(\wedge {\bf Q})$, we have also $Ar \in {\rm ran}(\wedge {\bf Q})$.  This
is equivalent to the statement
\[
(\wedge {\bf Q})A(\wedge {\bf Q}) = A(\wedge {\bf Q}) .  
\]

Next, note that for any projection $P$, $[A, P] = 0$
if and only if $[A^{\dagger}, P] = 0$.  So by repeating the above argument
with $A^{\dagger}$ in place of $A$, we find 
\[
(\wedge {\bf Q})A^{\dagger}(\wedge {\bf Q}) = A^{\dagger}(\wedge {\bf Q}).
\]  
Taking adjoints, this becomes
\[
(\wedge {\bf Q})A(\wedge {\bf Q}) = (\wedge {\bf Q})A.
\]  
Comparing with the earlier result $(\wedge {\bf Q})A(\wedge {\bf Q}) =
A(\wedge {\bf Q})$, we conclude that $(\wedge {\bf Q})A = A(\wedge {\bf Q})$,
and the claim is proved.

The claim shows that given a set of projections ${\bf Q}$, the meet of its
elements, $\wedge {\bf Q}$, commutes with any operator that commutes with
every projection in $\bf Q$.  In other words, $\wedge {\bf Q} \in {\bf Q}''$.
But if $\bf Q \subseteq \underline{{\bf P}^{\prime}} \subseteq {\bf
P}^{\prime}$, then ${\bf P}^{\prime \prime} \subseteq {\bf Q}^{\prime}$ which
in turn implies ${\bf Q}^{\prime \prime} \subseteq {\bf P}^{\prime \prime
\prime} = {\bf P}^{\prime}$.  So $\wedge {\bf Q} \in \underline{{\bf
P}^{\prime}}$ and $\underline{{\bf P}^{\prime}}$ is closed under taking
arbitrary meets of its elements.

An argument similar to the above establishes that if an operator commutes
with every projection in $\bf Q$, then it commutes with their join $\vee {\bf
Q}$.  Hence $\underline{{\bf P}^{\prime}}$ is closed under arbitrary joins
and is a complete lattice.

The rest is trivial.  Clearly $\underline{{\bf P}^{\prime}}$  contains 1 and
0, and if $P$ is in $\underline{{\bf P}^{\prime}}$ then so is $1 - P$.  So
the  orthocomplement on $\underline{{\bf P}^{\prime}}$ is just the
restriction of the orthocomplement operation on $L({\sf H})$, and \emph{ipso
facto} satisfies the orthomodular identity.  \ \emph{QED}.\\[\baselineskip] \indent
Generally, sublattices of $L({\sf H})$ need not be atomic if ${\sf H}$ is
infinite-dimen\-sional -- just consider the Boolean algebra generated by the
spectral projections of an observable with a continuous spectrum.  Under what
circumstances, then, can we be assured that a sublattice ${\bf d}$ with
$^{\ast}$-closed extension will be atomic?  The following corollary offers a
sufficient condition.
\begin{itemize}
\item  \emph{Discrete operator}.  Call a self-adjoint operator \emph{discrete}
if there exists $\varepsilon > 0$ such that no two elements of its spectrum
are closer than $\varepsilon$ to one another.
\end{itemize}
Then we have the following:\\[\baselineskip] 
{\bf Corollary}.  Given a subset
of projections ${\bf d} \subseteq L({\sf H})$, if $\overline{{\bf d}}$ is
$^{\ast}$-closed and contains only discrete observables, then ${\bf d}$ is
atomic.\\[\baselineskip]
\emph{Idea behind proof}:  If ${\bf d}$ is \emph{not} atomic, then we show by
explicit construction that $\overline{{\bf d}}$ contains a non-discrete
observable.\\[\baselineskip] {\bf Proof}.  (In the following, the indices $n$
and $N$ run over the positive integers 1, 2, 3, $\ldots$)

\emph{Step 1}.  In any non-atomic lattice $L$, there is a countable family of
distinct elements $\{x_{n}\} \subseteq L$ for which
\[
0 < \ldots < x_{3} < x_{2} < x_{1}.
\]

\emph{Proof}.  Observe first that if every non-zero, non-atomic element of $L$
contained an atom, then in fact every non-zero element of $L$ would contain
an atom, and $L$ would be atomic.  So if $L$ is non-atomic, then there must
be a non-zero, non-atomic element $x_{1}$ which does not contain an atom.

Since this element $x_{1}$ is non-zero and non-atomic, there must be a
non-zero element $x_{2}$ distinct from $x_{1}$ with
\[
0 < x_{2} < x_{1}.
\]
But $x_{1}$ does not contain an atom; so $x_{2}$ cannot be an atom; so
$x_{2}$ must contain a non-zero element $x_{3}$ distinct from $x_{2}$:
\[ 
0 < x_{3} < x_{2} < x_{1}.
\]
Accordingly, it is clear that whenever a non-atomic lattice $L$ contains the
distinct non-zero elements $x_{n} < \ldots < x_{2} < x_{1}$, where $x_{1}$
does not contain an atom, then it also contains a non-zero element $x_{n+1}$,
distinct from $x_{n}$, with $x_{n+1} < x_{n} < \ldots < x_{2} < x_{1}$.
Therefore, as $L$ does in fact contain such an element $x_{1}$, Step 1
follows by induction.

Applying Step 1 to the lattice ${\bf d}$, we have
\[ 
0 < \ldots < P_{3} < P_{2} < P_{1}
\]
for some family of distinct projections $\{ P_{n} \}
\subseteq {\bf d}$.

\emph{Step 2}.  There is a family of mutually orthogonal non-zero
projections $\{ M_{n} \} \subseteq {\bf d}$ and a projection $P_{\infty} \in
{\bf d}$ which, together with $P^{\perp}_{1} \in {\bf d}$,
form a mutually orthogonal, complete set, that is:
\[
\sum_{n=1}^{\infty} M_{n} + P_{\infty} + P^{\perp}_{1} =1.
\]
(The limit in the sum is a strong limit.)

\emph{Proof}.  First we define $P_{\infty}$.  Since (by the theorem) ${\bf d}$
is a complete lattice, the family of projections $\{ P_{n} \} \subseteq {\bf
d}$ has an infimum $\wedge \{ P_{n} \} \equiv P_{\infty} \in {\bf d}$.
(Alternatively, it is not hard to see that the infimum $P_{\infty}$ is none other than
the strong limit of the sequence $\{ P_{n} \}$:
\[
\lim_{n\rightarrow \infty} P_{n} = P_{\infty} .
\]
Thus, by $^{\ast}$-closure, $P_{\infty}$ is in $\overline{{\bf d}}$, hence in ${\bf d}$.)

Next, define the mutually orthogonal projections
\begin{eqnarray*}
M_{1} & = & P_{1} \wedge P_{2}^{\perp} = P_{1} - P_{2} \\ 
M_{2} & = & P_{2} \wedge P_{3}^{\perp} = P_{2} - P_{3} \\
&\vdots & \\
M_{n} & = & P_{n} \wedge P_{n+1}^{\perp} = P_{n} - P_{n+1} \\
&\vdots &
\end{eqnarray*}
(Since ${\bf d}$ is an ortholattice, each $M_{n}$ is in ${\bf d}$.)  Then
\begin{eqnarray*}
\sum_{n=1}^{\infty} M_{n} & = & \lim_{n\rightarrow \infty} (M_{1} + M_{2} +
\cdots + M_{n}) \\ 
& = & \lim_{n\rightarrow \infty} (P_{1} - P_{2}  + P_{2} - P_{3}  +
\cdots + P_{n} - P_{n+1} ) \\ 
& = & \lim_{n\rightarrow \infty} (P_{1} - P_{n+1} ) \\ 
& = & P_{1} - \lim_{n\rightarrow \infty} P_{n+1} \\ 
& = & P_{1} - P_{\infty} .
\end{eqnarray*}
So $\sum_{n=1}^{\infty} M_{n}$ exists and satisfies
\[
\sum_{n=1}^{\infty} M_{n} + P_{\infty} + P^{\perp}_{1} =1
\]
as claimed.  (Note $P^{\perp}_{1}$ is in ${\bf d}$ since ${\bf d}$ is an
ortholattice.)

Finally, note that since $P_{\infty} \leq P_{n}$ for all $n$, $P_{\infty}$ is
orthogonal to each $M_{n}$: $P_{\infty} M_{n} = P_{\infty} (P_{n} -  P_{n+1})
= P_{\infty} - P_{\infty} = 0$.  Thus one sees that
$\{ M_{n} \}$, $P_{\infty}$, and $P_{1}^{\perp}$ form a mutually orthogonal,
complete set within ${\bf d}$, as
claimed.

\emph{Step 3}.  There is a non-discrete observable $Q \in \overline{{\bf d}}$.

\emph{Proof}.  For each $N$, define
\[
Q_{N} = P_{1}^{\perp} +  e^{-1} M_{1}  +  e^{-2} M_{2}  +  \ldots  +  e^{-N}
M_{N}. 
\]
Let $v$ be a non-zero but otherwise arbitrary vector in the Hilbert space,
and consider the sequence $\{ Q_{N} v \}$.  It is elementary to show that
this is a Cauchy sequence.  (Hint:  Given $\varepsilon > 0$, let
$N_{\varepsilon}$ be greater than $\log (2 \| v \| / \varepsilon)$; note
also that $\| M_{n} v \| \leq \| v \|$.)  By completeness of the Hilbert
space, $\{ Q_{N} v \}$ therefore converges in norm.  Denoting the limit
vector by $q(v)$, we have then
\[
\lim_{N\rightarrow \infty} \| Q_{N} v - q(v) \| = 0
\]
whenever $v \neq 0$.  Further, if $v = 0$, then obviously
\[
\lim_{N\rightarrow \infty} \| Q_{N} v - 0 \| = 0
\]
so define $q(0) = 0$.  Then for each $v$ we have shown that
there is a $q(v)$
such that
\[
\lim_{N\rightarrow \infty} \| Q_{N} v - q(v) \| = 0 .
\]
By definition therefore the map $v \stackrel{Q}{\mapsto} q (v)$ defines the
strong limit operator $Q = \lim_{n\rightarrow \infty} Q_{n}$.  So we may
write
\[
Q = P_{1}^{\perp} + \sum_{n=1}^{\infty} e^{-n} M_{n} .
\]
Clearly this operator has spectrum $\{1, e^{-1}, e^{-2}, \ldots , e^{-n},
\ldots , 0 \}$ (with $Qv = 0$ for $v \in {\rm ran} (P_{\infty}))$, so it is
non-discrete.  Yet its spectral projections $P_{1}^{\perp}$, $\{ M_{n} \}$,
and $P_{\infty}$
are in ${\bf d}$; so $Q$ is in $\overline{{\bf d}}$.  This establishes Step 3, and the
corollary is proved. \ \emph{QED}.\\[\baselineskip]

It should be noted that the converse to Thm.~3 fails.  Consider a
two-dimen\-sional Hilbert space ${\sf H}_{2}$ and take the (trivially)
complete, atomic and orthomodular lattice of projections ${\bf d}$ generated
by two distinct pairs of orthogonal, one-dimensional projections in the
plane.  There can be no ${\bf P}$ satisfying ${\bf d} = \underline{{\bf
P}^{\prime}}$ in this case.  For since ${\bf d} \neq L({\sf H}_{2})$, ${\bf P}$
must contain something other than 0 or 1.  So ${\bf P}$ must contain a
one-dimensional projection.  But there is no such projection that all four
one-dimensional projections in ${\bf d}$ commute with.  \mbox{In fact,} it is easy
to see that a (proper) subortholattice of $L({\sf H}_{2})$ extends \mbox{to a
$^{\ast}$-closed} set of observables only if it is a Boolean algebra (i.e.\
distributive ortholattice).

\section{Valuations on Sets of Definite-Valued Observables}

\subsection{Homomorphisms and Valuations}

To this point we have been focusing on \emph{structural} questions regarding
the set of definite-valued operators.  The subject one might naturally wish
to consider next is that of \emph{value assignments} on the set of definite-valued
operators.  After all, for such a set to be dubbed `definite-valued,' it must
admit valuations!  In this section we analyze valuations from a structural
perspective.

As usual, we will need to introduce some definitions.
\begin{itemize}
\item \emph{Two-Valued (Ortholattice) Homomorphism}.  Given an ortholattice
$L$, a \emph{two-valued ortholattice homomorphism} is a map $[.] : L
\rightarrow \{ 0,1 \} \subset \mathbb{R}$ which respects the operations of orthocomplement, meet
and join:
\begin{eqnarray*}
[ x^{\perp} ] & = & 1 - [x] \\
\mbox{} [ x \wedge y ] & = & [x] \cdot [y] \\
\mbox{} [ x \vee y ] & = & [x] + [y] - [x] \cdot [y] .
\end{eqnarray*}
(The right-hand sides of these equations are arithmetic operations
in $\mathbb{R}$ involving the numbers 0 and 1.)
\item \emph{Faithful Valuation}.  Consider a set ${\bf D}$ of self-adjoint
operators with polynomial $^{\ast}$-closure.  We use the term \emph{faithful
valuation} to refer to a real-valued map $\langle .\rangle  : {\bf D}
\rightarrow \mathbb{R}$ which assigns to each operator $Q$ a value in its
spectrum, and which satisfies
\begin{eqnarray*}
\langle a Q + S \rangle & = & a\langle Q \rangle  + \langle S \rangle \\ 
\langle Q^{2} \rangle & = & \langle Q \rangle^{2}.
\end{eqnarray*}
(Here $a$ is any real scalar.)
\item \emph{Functional Valuation}.  Consider a set ${\bf D}$ of self-adjoint
operators with $^{\ast}$-closure.  We use the term \emph{functional valuation}
to refer to a faithful valuation $\langle .\rangle  : {\bf D}
\rightarrow \mathbb{R}$ which satisfies
\[
\lim_{n \rightarrow \infty} \langle F_{n} \rangle = \langle F \rangle
\]
whenever the sequence $\{ F_{n} \} \subseteq {\bf D}$ converges strongly to
$F \in {\bf D}$.
\end{itemize}

A faithful valuation respects the \emph{polynomial} relationships among the
operators in a set with polynomial $^{\ast}$-closure.  To be precise, if
$F(Q_{1} , \ldots , Q_{k})  \in {\bf D}$ is a polynomial function of some
operators $Q_{i} \in {\bf D}$, then consider the natural corresponding 
\emph{real-valued} polynomial $f(x_{1},...,x_{k})$, where each $x_{i}$ takes
on values from the spectrum of $Q_{i}$.  In this case, a faithful valuation
$\langle .\rangle$ will satisfy
\[
\langle F(Q_{1} , \ldots , Q_{k}) \rangle  = f( \langle Q_{1} \rangle ,
\ldots ,\langle Q_{k} \rangle  ) .
\]

A \emph{functional} valuation, on the other hand, respects \emph{arbitrary}
functional relationships among the operators in a $^{\ast}$-closed set.  To
be precise, suppose that a sequence of polynomial functions $\{ F_{n}(Q_{1} ,
\ldots , Q_{k}) \}$ approaches an operator $F$ more and more closely in the
strong operator topology.  Then for a \emph{functional} valuation $\langle
.\rangle$  the numbers
\[
\langle F_{n} (Q_{1} , \ldots , Q_{k}) \rangle  = f_{n} ( \langle Q_{1}
\rangle , \ldots ,\langle Q_{k} \rangle  )
\]
must approach the number $\langle F\rangle$  more and more closely.  In other
words, the number assigned to the strong limit of a sequence $\{F_{n}\}$ is
the limit of the sequence of numbers assigned to the $F_{n}$'s.  And each of
these numbers is obtained in the natural way from the numbers assigned to the
$Q_{i}$.  This is what is meant by the phrase ``the mapping $\langle
. \rangle$  respects arbitrary functional relationships.''

It is worth emphasizing that ``arbitrary functional relationships'' means
\emph{arbitrary} functional relationships.  To say that $F$ is a function of
operators $Q_{1}$, $\ldots$, $Q_{k}$ means no more than that $F$ is the limit
of a sequence of polynomial functions of the $Q_{i}$.  According to this
definition, the operator $F$ need not be representable as a series expansion
in the $Q_{i}$, nor need it be in any sense a `continuous' function of the
$Q_{i}$.

\subsection{QuasiBoolean Algebras and Homomorphisms}

We are aiming to analyze the requirement that a $^{\ast}$-closed set of
observables admit enough functional valuations that the statistics prescribed
by quantum mechanics for observables within the set can be represented as
measures over the set of functional valuations on the set.  In this section
we lay the groundwork for showing that whether or not this requirement can be
met has everything to do with whether the projections in the $^{\ast}$-closed
set form a certain kind of ortholattice, dubbed a quasiBoolean algebra by
Bell and Clifton [1995].  Just as von Neumann algebras capture the structure
required of a $^{\ast}$-closed set for it to be functionally closed,
quasiBoolean algebras capture the structure required of the \emph{projections}
in a $^{\ast}$-closed set in order for it to admit enough functional
valuations to satisfy quantum statistics.

Here is one last round of definitions leading up to the
concept of a
quasiBoolean algebra.
\begin{itemize}
\item \emph{Ideal}.  An \emph{ideal} $I$ of a lattice $L$ is a (non-empty)
subset of $L$ such that:
\begin{equation}
\nonumber
\begin{aligned}
\mbox{} & x \in I, y \leq x \; \: \Rightarrow \; \: y \in I \\ 
\mbox{} & x,y \in I \; \: \Rightarrow \; \: x \vee  y \in I \\
\mbox{} & \qquad \quad 1 \not\in I.
\end{aligned}
\end{equation}
\item \emph{Principal ideal}.  For any $x \neq 1$ in a lattice $L$, the
set $x \! \! \downarrow \, \, \equiv \{ y \in L: y \leq
x \}$ is an ideal, called the \emph{principal ideal generated by} $x$.
\item $I$-\emph{quasiBoolean algebra}.  An ortholattice $L$ containing an
ideal $I$ is called an $I$-quasidistributive ortholattice, or an 
$I$-\emph{quasiBoolean algebra}, if for any $x \not\in I$ there is a
two-valued (ortholattice) homomorphism $[.] : L \rightarrow \{ 0 , 1 \}$ for
which $[x] = 1$.

(See Bell and Clifton [1995].  The terminology derives from the fact that
distributive ortholattices, i.e.\ Boolean algebras, satisfy the stronger
condition that for any $x \neq y$ there is a two-valued homomorphism $[.] :
L\rightarrow \{ 0,1 \}$ for which $[x] \neq [y]$.)
\end{itemize}

How does the abstract lattice-theoretic concept of a quasiBoolean algebra
connect with our problem?  Well, we are interested in characterizing
$^{\ast}$-closed sets of observables that support enough functional
valuations to satisfy quantum statistics.  Imagine then that we have a
$^{\ast}$-closed set ${\bf D}(W)$, and we want to know if it fits the bill.
Considering only the projections in ${\bf D}$, notice that we would certainly
run into trouble if there were some projection $P$ in ${\bf D}$ for which
${\rm Prob}_{W}(P=1) \neq 0$, but for which no functional valuation on ${\bf
D}$ allowed $P$ to take the value 1.  For then the measure of the set of
functional valuations sending $P$ to 1 would have to be zero, and our
hidden-variable theory would be doomed to `statistical failure.'

Now let's put the same argument somewhat differently.  Suppose that
$\underline{{\bf D}}$ 
does \emph{not} form an $I$-quasiBoolean algebra with
respect to the ideal 
$I = \{ P \in \underline{{\bf D}}: {\rm Prob}_{W} (P=1) =
0\}$. Then there is some projection $P$ with ${\rm Prob}_{W}(P=1) \neq 0$ for
which no homomorphism assigns $P$ the value 1.  Now, it is intuitively
plausible that if there were a functional valuation on ${\bf D}$ sending $P$
to 1, then by considering the restriction, there would also be a homomorphism
on the underlying ortholattice $\underline{{\bf D}}$ sending $P$ to 1.  (We
shall be addressing this and related issues in section 3.3 below.)  Taking
this on faith for the time being, then since in our scenario there is no
homomorphism sending $P$ to 1, there could not be a functional valuation
sending $P$ to 1.  Hence the measure of the functional valuations sending $P$
to 1 would have to be zero, in conflict with the quantum mechanical
probability ${\rm Prob}_{W}(P=1) \neq 0$.  In short, a functionally closed
modal interpretation is doomed to statistical failure unless $\underline{{\bf
D}}$ forms an $I$-quasiBoolean algebra with respect to the ideal $I = 
\{ P \in \underline{{\bf D}}:
{\rm Prob}_{W}(P=1) = 0 \}$.  Hence a mathematically appropriate object to seek for a
modal interpretation that \emph{does} satisfy quantum statistics is a
$^{\ast}$-closed set ${\bf D}$ whose projections form an $I$-quasiBoolean
algebra with respect to the ideal  $I = \{ P \in \underline{{\bf D}}: {\rm
Prob}_{W}(P=1) = 0 \}$.

As we have said, the primary aim of this paper is to describe a sense in
which $I$-quasiBoolean algebras are both necessary and sufficient to generate
functionally closed modal interpretations with enough functional valuations
to satisfy quantum statistics.  But before we can reach the goal, we need a
tractable characterization of their lattice structure.  (This will make it
easy to check that, for example, projection sets of ${\bf X}$-form have the
required properties.)  At present we only have a clean characterization for
complete, orthomodular, \emph{atomic} $I$-quasiBoolean algebras of
projections, so we are forced to depend on the Corollary to Thm.~3; and in
its present form, this Corollary dictates that we confine the rest of our
results to $^{\ast}$-closed sets of \emph{discrete} observables.  This does
not mean that we are specializing to the case of finite-dimensional Hilbert
spaces; but it does mean that from this point forward our results are only
complete for that case.

Since our characterization of $I$-quasiBoolean algebras does not make any use
of Hilbert space beyond its ortholattice structure, we shall present a purely
lattice-theoretic result (which extends the results of Bell and Clifton
[1995]).
\\[\baselineskip]
{\bf Theorem 4}.  Let $L$ be a complete, atomic, orthomodular lattice with
ideal $I$.  Then $L$ is an $I$-quasiBoolean algebra if and only if there is a
non-empty, mutually orthogonal subset $A$ of $L$, not containing 0, such
that:
\renewcommand{\labelenumi}{(\theenumi).}
\begin{enumerate}
\item For any $y \in L$, $a \leq y$ or $a \leq y^{\perp}$ for all $a \in A$;
and 
\item $I = (\vee A)^{\perp}\! \! \downarrow$.
\end{enumerate}
{\bf Proof}. ($\Leftarrow$)  Let $y \in L$ be such that $y \not\in  I = (\vee
A)^{\perp}\! \! \downarrow$ (by (2)).  We must show that there is a two-valued
homomorphism on $L$ sending $y$ to 1.  Since $y \not\in (\vee A)^{\perp}
\! \! \downarrow$, there must be an element $b \in A$ such that $b \leq y$.  (For
if not, then by (1) $a \leq y^{\perp}$ for all $a \in A$, which implies $\vee
A \leq y^{\perp}$, i.e.\ $y \leq (\vee A)^{\perp}$.  But then $y \in (\vee
A)^{\perp}\! \! \downarrow$, contradicting our hypothesis.) Invoking (1) (together
with the fact that $b \neq 0$), construct the well-defined mapping $[.]^{b} :
L \rightarrow
\{0,1\}$ by
\begin{eqnarray*}
[x]^{b} = 1 \; & {\rm if} \; & b \leq x; \\
\mbox{} [x]^{b} = 0 \; & {\rm if} \; & b \leq x^{\perp}.
\end{eqnarray*}
By definition then, $[y]^{b} = 1$.  To complete the argument we verify that
$[.]^{b}$ is an ortholattice homomorphism.  First, observe that $[x]^{b} = 1$
if and only if $[x^{\perp}]^{b} = 0$.  Next, for any $x_{1},x_{2} \in L$, we
have
\begin{eqnarray*}
[x_{1} \wedge x_{2}]^{b} = 1 & \Leftrightarrow &  b \leq x_{1} \wedge x_{2} \\
& \Leftrightarrow & b \leq x_{1} \; {\rm and} \; b \leq x_{2} \\
& \Leftrightarrow & [x_{1}]^{b} = [x_{2}]^{b} = 1 \\
& \Leftrightarrow & [x_{1}]^{b} \cdot [x_{2}]^{b} = 1.
\end{eqnarray*}
Thus $[x_{1} \wedge x_{2}]^{b} = [x_{1}]^{b} \cdot [x_{2}]^{b}$.  Lastly,
since $L$ is an ortholattice it is subject to de Morgan's laws, 
so the equation $[x_{1}
\vee x_{2}]^{b} = [x_{1}]^{b} + [x_{2}]^{b} - [x_{1}]^{b} \cdot [x_{2}]^{b}$
follows from preservation of orthocomplement and meet under the 
\mbox{mapping $[.]^{b}$.}

($\Rightarrow$)  Now suppose $L$ is an $I$-quasiBoolean algebra.  Let $A$ be
the set of all atoms in $L$ not contained in $I$ (so that $A \cap I =
\varnothing $).  (Note that $A$ cannot contain 0 since it contains only
atoms.  And $A$ is non-empty, for otherwise every atom contained in 1 would
lie in I; and since 1 is the join of its atoms, this would force the
contradiction $1 \in I$.)  For the proof of (1) suppose, for \emph{reductio
ad absurdum}, that there are $y \in L$ and $b \in A$ such that $b \not\leq y$
and $b \not\leq y^{\perp}$.  Since $L$ is $I$-quasiBoolean and $A \cap I =
\varnothing$, there is a two-valued homomorphism $[.] : L\rightarrow \{ 0,1
\}$ sending $b$ to 1.  Since $b$ is an atom, $b \wedge y = b \wedge y^{\perp}
= 0$.  Therefore, $[y] = [b] \cdot [y] = [b\wedge y] = 0$, and by the same
token $[y^{\perp}] = 0$, which is a contradiction.  (It now follows from (1)
that the elements of $A$ must be mutually orthogonal.)  To show (2), note
that all atoms in $(\vee A)^{\perp}$ are in $I$ (otherwise, by the definition
of $A$ there would be an element $b \in A$ such that $b \leq (\vee
A)^{\perp}$, implying $b \leq b^{\perp}$ and hence the contradiction $b =
0$).  Since $(\vee A)^{\perp}$ is the join of its atoms and $I$ is an ideal,
$(\vee A)^{\perp} \in I$ which implies $(\vee A)^{\perp}\! \! \downarrow \,
\, \subseteq I$.  For equality, suppose that for some $y \in I$, $y \not\in
(\vee A)^{\perp} \! \! \downarrow$; that is, $y \not\leq (\vee A)^{\perp}$.
By (1) (just proved) there must be an element $b \in A$ such that $b \leq y$.
But then since $y \in I$, $b \in I$ contradicting $A \cap I = \varnothing$.
Thus $I = (\vee A)^{\perp}\! \! \downarrow$. \ \emph{QED}.\\[\baselineskip]
\indent Returning now to our favorite example, sets of ${\bf X}$-form, we get
what we were after:\\[\baselineskip] 
{\bf Corollary}.  If a set of projections ${\bf d}$ is of ${\bf X}$-form,
then it is an $I$-quasiBoolean algebra where
\[
I = \{ P \in {\bf d}: P \sum_{X\in {\bf X}} X = 0 \} .
\]  
{\bf Proof}.  The
Corollary to Thm.~2 establishes that $\overline{{\bf d}}$ is $^{\ast}$-closed.
So Thm.~3
establishes that ${\bf d}$ is a complete orthomodular lattice.  Since ${\bf
X}$-form lattices are clearly atomic, with the atoms being the $X\in {\bf X}$
and all one-dimensional projections orthogonal to all the $X\in {\bf X}$, the
conclusion follows immediately from Thm.~4 (with ${\bf X}$ playing the role
of $A$).  \ \emph{QED}.

\subsection{Projections and Functional Valuations}

We are now in a position to fill in the last piece of our puzzle before
taking a look at exactly how these technical results sidestep von Neumann's
no-hidden-variables theorem.  Our final theorem simplifies the task of
deciding whether a given $^{\ast}$-closed set will support enough functional
valuations to satisfy quantum statistics, by substituting the simpler
question of whether its underlying set of projections forms the appropriate
quasiBoolean algebra.\\[\baselineskip]
{\bf Theorem 5}.  Let ${\bf d}$ be a set of projections with $\overline{{\bf d}}$ a
$^{\ast}$-closed set of definite-valued operators having discrete spectra,
and let $W$ be a density operator.  Then the following are equivalent:
\begin{enumerate}
\item There is a probability measure $\mu$ on the set of all functional
valuations $\langle .\rangle  : \overline{{\bf d}} \rightarrow \mathbb{R}$ such
that for any mutually commuting subset $\{A , B , C , \ldots \}$ of
$\overline{{\bf d}}$ and corresponding sets of eigenvalues $\{ \alpha , \beta
, \gamma , \ldots \}$:
\[ 
{\rm Prob}_{W} (A \in \alpha, B \in \beta, C \in \gamma,...) = \mu \{
\langle .\rangle\! : \langle A \rangle \in \alpha, \langle B\rangle \in \beta,
\langle C \rangle \in \gamma,...\}.\] 

\item ${\bf d}$ is an $I$-quasiBoolean algebra, where $I = \{ P \in {\bf d}:
PW = 0 \}$.
\end{enumerate}
{\bf Proof}.

\underline{(1) $\Rightarrow$ (2)}

Since $\overline{{\bf d}}$ is $^{\ast}$-closed, ${\bf d}$ is a complete ortholattice (by
Thm.~3).  Assuming the existence of a probability measure $\mu$ satisfying
(1), we must show that ${\bf d}$ forms an $I$-quasiBoolean algebra.  So let
$P$ be any element of ${\bf d}$ such that $PW \neq 0$ (i.e.\ $P \not\in  I$).
Then $P$ is in $\overline{{\bf d}}$ and so, by (1), there exists a probability measure
$\mu$ such that:
\[
{\rm Prob}_{W} (P=1) = \mu \{ \langle .\rangle : \langle P \rangle =1\} .
\]
But since $PW \neq 0$, ${\rm Prob}_{W} (P=1) = {\rm Tr}(PW) \neq 0$,
therefore $\mu \{ \langle .\rangle : \langle P\rangle =1 \} \neq 0$.  So
there exists a functional valuation on $\overline{{\bf d}}$ sending $P$ to $1$.  Since we
seek a \emph{homomorphism} sending $P$ to $1$, it suffices to complete the
proof if we can show that every functional valuation on $\overline{{\bf d}}$ restricts to
an ortholattice homomorphism on ${\bf d}$.

Let $\langle .\rangle : \overline{{\bf d}} \rightarrow \mathbb{R}$ be a functional valuation.
Consider a projection $P$ in ${\bf d}$ and its complement $P^{\perp} \in {\bf
d}$.  Then $P$, $P^{\perp} \in \overline{{\bf d}}$  satisfy $P + P^{\perp} = 1$, so from
$\langle aQ+S \rangle  = a\langle Q\rangle  + \langle S\rangle$  we have
$\langle P\rangle  + \langle P^{\perp}\rangle  = \langle 1\rangle  = 1$, or
\[
\langle P^{\perp}\rangle  = 1 - \langle P\rangle .
\]

Next, let $P_{1}$ and $P_{2}$ be two projections in ${\bf d}$, with $P_{1}
\wedge P_{2} \in {\bf d}$ their meet.  It is easily verified that $P_{1}
\wedge P_{2} = \lim_{n \rightarrow \infty} (\frac{1}{2} [ P_{1} P_{2} + P_{2}
P_{1}])^{n}$, and both $P_{1} \wedge P_{2}$ and $(\frac{1}{2} [ P_{1} P_{2} +
P_{2} P_{1}])^{n}$ lie in $\overline{{\bf d}}$ (by $^{\ast}$-closure).  So by
functionality of $\langle .\rangle$ we must have
\[
\langle P_{1} \wedge P_{2} \rangle = \lim_{n \rightarrow \infty} \langle
(\textstyle{\frac{1}{2}} [ P_{1} P_{2} + P_{2} P_{1}])^{n} \rangle.
\]
Now, by faithfulness, $\langle (\frac{1}{2} [ P_{1} P_{2} + P_{2} P_{1}])^{n}
\rangle = \langle \frac{1}{2} [ P_{1} P_{2} + P_{2} P_{1}] \rangle^{n}$. And
again by faithfulness, for any $Q,S \in \overline{{\bf d}}$ we have $\langle
\frac{1}{2} (QS+SQ) \rangle  = \langle Q\rangle \cdot \langle S\rangle$. (For
the proof, use $\frac{1}{2} (QS+SQ) = \frac{1}{4} (Q+S)^{2} - \frac{1}{4}
(Q-S)^{2}$, and note that $\langle 0\rangle  = 0$.) Thus $\langle \frac{1}{2}
[ P_{1} P_{2} + P_{2} P_{1}] \rangle^{n} = (\langle P_{1} \rangle \cdot
\langle P_{2} \rangle )^{n} = \langle P_{1} \rangle^{n} \langle P_{2}
\rangle^{n}$.  But since $\langle . \rangle$  is a valuation, it assigns to
$P_{1}$ and $P_{2}$ the values 0 or 1, so in either case $\langle P_{i}
\rangle^{n} = \langle P_{i} \rangle$.  Hence $\langle P_{1} \rangle^{n}
\langle P_{2} \rangle^{n} = \langle P_{1} \rangle \cdot \langle P_{2}
\rangle$  for each $n$, and so we have
\[
\langle P_{1} \wedge P_{2} \rangle  = \langle P_{1} \rangle \cdot \langle
P_{2} \rangle .
\]

Finally, $\langle P_{1} \vee P_{2} \rangle  = \langle P_{1} \rangle  +
\langle P_{2} \rangle - \langle P_{1} \rangle \cdot \langle P_{2} \rangle$
follows by de Morgan's law.  So we have established that $\langle .\rangle$
restricted to ${\bf d}$ is an ortholattice homomorphism.  This completes the
proof that ${\bf d}$ is an $I$-quasiBoolean algebra with respect to $I = \{ P
\in {\bf d}: PW = 0 \}$.\\[\baselineskip] 
\indent \underline{(2) $\Rightarrow$ (1)}

Now suppose ${\bf d}$ is $I$-quasiBoolean, where $I = \{ P \in {\bf d}: PW =
0 \}$.  We must exhibit a probability measure $\mu$ satisfying (1).  As
discussed earlier, a \emph{necessary} condition for the existence of such a
$\mu$ is that the following claim hold:

\underline{Claim}: For any $P \in {\bf d}$ such that ${\rm Prob}_{W}(P=1)
\neq 0$, there exists a functional valuation $\langle .\rangle  :
\overline{{\bf d}} \rightarrow \mathbb{R}$ sending $P$ to 1.

To establish that this is in fact the case, we make use of Thm.~4 and the
corollary to Thm.~3.  According to these results, since ${\bf d}$ is an
$I$-quasiBoolean algebra of projections with a $^{\ast}$-closed extension,
and since ${\bf d}$ is assumed to contain only discrete observables, it
follows that there is a set of mutually orthogonal projections ${\bf X}
\subseteq {\bf d}$ such that:
\[
{\bf d} \subseteq \{ P: PX = X \; {\rm or} \; 0 \; \mbox{for all} \; X \in
{\bf X} \}, 
\]
\[
I = \{ P \in {\bf d}: P \sum_{X \in {\bf X}} X = 0 \} = \{ P \in {\bf d}: PX
= 0 \; \mbox{for all} \; X \in {\bf X} \}.
\]
Since $I = \{ P \in {\bf d}: PW = 0 \}$, it follows that for $P \in {\bf d}$, $PW
\neq 0$ is equivalent to $PX \neq 0$ for some $X \in {\bf X}$, which is in
turn equivalent to $PY = Y$ for some $Y \in {\bf X}$.

Now consider any $R \in {\bf d}$ such that ${\rm Prob}_{W} (R=1) = {\rm
Tr}(RW) \neq 0$.  Then $RW \neq 0$, so $RY = Y$ for some $Y \in {\bf X}$.
The mapping $[.] : {\bf d} \rightarrow \{ 0,1 \}$ given by
\begin{eqnarray*}
[P] = 1 \; &{\rm if}& \; PY = Y \\
\mbox{} [P] = 0 \; &{\rm if}& \; PY = 0
\end{eqnarray*}
is easily verified (as in the first part of Thm.~4) to be an
ortholattice
homomorphism which sends both $R$ and $Y$ to 1.  So, to complete
the proof of
the claim, we need to show that the homomorphism $[.]$ on ${\bf d}$
extends to a
functional valuation $\langle .\rangle$  on $\overline{{\bf d}}$.  (For this we will
eventually have to recall
that $[.]$ has been defined so that $[Y] = 1$, and that $Y \in {\bf 
X}$.)

Define a map $\langle .\rangle  : \overline{{\bf d}} \rightarrow \mathbb{R}$ as follows.  For
an operator $Q \in \overline{{\bf d}}$, let $Q = \sum q_{i} Q_{i}$ be its spectral
resolution (remember $\overline{{\bf d}}$ consists of only discrete spectra observables),
and define
\[
\langle Q \rangle \equiv  \sum q_{i} [Q_{i}] .
\]
It is clear that $\langle .\rangle$  agrees with $[.]$ on ${\bf d}$, since
for a projection $P \in {\bf d}$ we have $\langle P\rangle  \equiv \sum p_{i}
[P_{i}] =1 \cdot [P]$.

We argue next that $\langle .\rangle$  is a faithful valuation on 
$\overline{{\bf d}}$.

First of all, since $\sum Q_{i} = 1$, it is easy to show that $[.]$ must
assign the value 1 to exactly one of the projections $Q_{i}$.  One sees
therefore that $\langle . \rangle$ assigns to $Q$ a value in its spectrum.

Second, $\langle .\rangle$  has the property that $\langle aQ+S\rangle  = a\langle Q\rangle 
+ \langle S\rangle$. 
To
see this, let $C
= aQ + S$, which, phrased in terms of spectral resolutions,
reads
\[
\sum c_{i} C_{i} = a \sum q_{j} Q_{j} + \sum s_{k} S_{k}.
\]
Since $[.]$ is an ortholattice homomorphism, there exist unique
$i'$, $j'$, and
$k'$ such that $[C_{i'}] = [Q_{j'}] = [S_{k'}] = 1$.  For these
projections we will
therefore have $[C_{i'} \wedge Q_{j'} \wedge S_{k'}] = 1\cdot 1 \cdot 1 = 1$.  It follows
that
$C_{i'} \wedge Q_{j'} \wedge S_{k'}$ is a
non-zero projection, hence there is a non-zero vector $v$ in the
range of
$C_{i'} \wedge Q_{j'} \wedge S_{k'}$.  Applying both sides of the 
above spectral
resolution equation to
this vector $v$, we find
\[
c_{i'} = aq_{j'} + s_{k'}.
\]
Since $\langle C\rangle$  is none other than the eigenvalue $c_{i'}$ for
which $[C_{i'}] = 1$, and similarly for $\langle Q\rangle$  and $\langle
S\rangle$, this just says that $\langle C\rangle = a\langle Q\rangle  +
\langle S\rangle $.  \mbox{Thus $\langle aQ+S\rangle  =$} $a\langle Q\rangle  +
\langle S\rangle $, as was to be shown.

Third, $\langle .\rangle$  has the property that $\langle Q^{2}\rangle  =
\langle Q\rangle^{2}$. To see this, let $C = Q^{2}$, which, phrased in terms
of spectral resolutions, reads
\[
\sum c_{i} C_{i} = \sum q_{j}^{2} Q_{j}.
\]
Imitating the above reasoning, we find
\[
c_{i'} = q_{j'}^{2}
\]
which says that $\langle C\rangle  = \langle Q\rangle^{2}$.  Thus $\langle
Q^{2}\rangle = \langle Q\rangle^{2}$, as was to be shown.

These three arguments establish that $\langle .\rangle  : \overline{{\bf d}} \rightarrow
\mathbb{R}$ is a \emph{faithful valuation}. We show next that $\langle .\rangle$ is a \emph{functional valuation}.  For let $Q_{1},
\ldots , Q_{k}$ be operators in $\overline{{\bf d}}$, with $\{ F_{n} (Q_{1}, \ldots ,
Q_{k}) \}$ a sequence of self-adjoint polynomials in the $Q_{i}$ converging
strongly to $F$.  Since $\overline{{\bf d}}$ is $^{\ast}$-closed, each $F_{n}$ belongs to
$\overline{{\bf d}}$ , and $F$ is in $\overline{{\bf d}}$ as well.  By definition then, in the
spectral resolutions $F_{n} = \sum f_{i}^{n} F_{i}^{n}$ and $F = \sum f_{j}
F_{j}$, the projections $F_{i}^{n}$ and $F_{j}$ are all in ${\bf d} \subseteq
\{ P: PX = X \; {\rm or} \; 0 \; $for all$ \; X \in {\bf X} \}$.
Therefore, since $Y$ is an element of ${\bf X}$, we have 
\[ 
F_{n} Y = \sum f_{i}^{n} F_{i}^{n} Y \equiv q_{n}Y
\]
where $q_{n}$ is a real scalar. Similarly
\[
F Y = \sum f_{j} F_{j} Y \equiv qY
\]
where $q$ is another real scalar.  Furthermore, from the fact that $\{ F_{n}
\} \rightarrow F$ strongly, it follows easily that $\{ q_{n} \} \rightarrow
q$ in modulus.  For let $w$ be a unit vector in the range of $Y$.  (Recall
that such a vector exists since $[Y] = 1$.)  Then we have
\begin{eqnarray*}
| q_{n}-q |  &=& \| (q_{n}-q)w \| \\
&=& \| (q_{n}-q)Yw \| \\
&=& \| (F_{n}-F)w \| \\
&\rightarrow&  0
\end{eqnarray*}
by strong convergence of $\{ F_{n} \}$ to $F$.

Next, from
\[
q_{n} \rightarrow q
\]
we have
\[
q_{n} \langle Y \rangle  \rightarrow q \langle Y \rangle
\] 
(recall that $\langle Y\rangle  = 1$)
\[
\Rightarrow \qquad \langle q_{n} Y\rangle  \rightarrow \langle q Y\rangle
\] 
(since $\langle .\rangle$  is faithful)
\begin{eqnarray*}
& \Rightarrow & \qquad \langle F_{n} Y \rangle \rightarrow  \langle F Y
\rangle \\ 
\mbox{} & \Rightarrow & \qquad \langle \textstyle{\frac{1}{2}} (F_{n} Y+Y
F_{n}) \rangle \rightarrow  \langle \textstyle{\frac{1}{2}} (FY+YF)\rangle
\end{eqnarray*}
(since $F_{n}$ and $F$ both commute with $Y$)
\[
\Rightarrow \qquad \langle F_{n} \rangle \cdot \langle Y\rangle  \rightarrow
\langle F\rangle \cdot \langle Y\rangle  
\]
(since $\langle \cdot\rangle$ is faithful).  But since $\langle Y\rangle  =
1$, this last statement requires
\[
\langle F_{n} \rangle  \rightarrow  \langle F\rangle ,
\]
so that $\langle .\rangle$ is functional as promised.  This finally
establishes the claim:  For any $P \in {\bf d}$ such that ${\rm
Prob}_{W}(P=1) \neq 0$, there exists a functional valuation $\langle .\rangle
: \overline{{\bf d}} \rightarrow \mathbb{R}$ sending $P$ to 1.

Having established this, we can now easily define a probability measure
satisfying (1) as follows.  Let our measure space consist of the set ${\cal
F}$ of all functional valuations on $\overline{{\bf d}}$; let the measurable sets ${\tt
M}$ be sets of the form $S_{P} = \{ \langle .\rangle \in {\cal F}: \langle
P\rangle =1\}$ for some $P$ in $\overline{{\bf d}}$; and let the measure be defined by
\[
\mu \{ \langle .\rangle \in {\cal F}: \langle P \rangle =1\} \equiv {\rm
Prob}_{W}(P=1).
\]

In order to show that everything is well-defined, we first show that $\langle
{\cal F},{\tt M},\mu \rangle$ is a probability space.

${\tt M}$ constitutes a sigma field on ${\cal F}$.  For $\varnothing  = S_{0}
\in {\tt M}$, ${\cal F} = S_{1} \in {\tt M}$, and $(S_{P})^{\rm c} = S_{1-P}
\in {\tt M}$.  Furthermore, $\cap_{i} S_{P_{i}} = S_{\wedge_{i}P_{i}} \in
{\tt M}$ since ${\bf d}$ is a complete lattice (Thm.~3), and
\begin{eqnarray*}
\wedge_{i}^{n} P_{i}  &\rightarrow &  \wedge_{i} P_{i} \\
\Rightarrow   \qquad \langle \wedge_{i}^{n} P_{i} \rangle & \rightarrow &
\langle \wedge_{i} P_{i}\rangle \\
\Rightarrow \qquad \Pi_{i}^{n} \langle P_{i} \rangle & \rightarrow & \langle
\wedge_{i} P_{i} \rangle 
\end{eqnarray*}
which implies that $\langle \wedge_{i} P_{i} \rangle  = 1$ exactly when
$\langle P_{i}\rangle = 1$ for all $i$.  It follows from de Morgan's law that
$\cup_{i}S_{P_{i}} = S_{\vee_{i} P_{i}} \in {\tt M}$.

The map $\mu$ is also a probability measure.  It takes values in the interval
$[0,1]$; it satisfies $\mu (\varnothing ) = 0$ (thanks to the claim); and it
satisfies $\mu ({\cal F}) = 1$.  To show that $\mu$ is countably additive,
suppose we have mutually disjoint $\{ S_{P_{i}} \}$, so the meet of any two
projections in the set $\{ P_{i} \} \subseteq {\bf d}$ is zero.  Recall that
since ${\bf d}$ is $I$-quasiBoolean, for all projections in ${\bf d}
\subseteq \{ P: PX = X$ or 0 for all $X \in {\bf X} \}$, $PX=0$ for all $X \in
{\bf X}$ if and only if $PW=0$.  This latter condition implies that the
ranges corresponding to the $X \in {\bf X}$ span the image space of $W$,
i.e.\ the subspace generated by its non-zero eigenvalue eigenspaces.  Now let
$x$ be a vector in the range of one of the $X \in {\bf X}$.  It follows that
$(\vee_{i} P_{i} ) x = (\sum_{i} P_{i} )x$.  For each $P_{i}$ must satisfy
$P_{i} x = x$ or 0; and since any two projections in the set $\{ P_{i} \}$
have meet 0, this implies that either $P_{i} x = 0$ for all $i$ (in which
case ($\vee_{i} P_{i})x = (\sum_{i} P_{i})x = 0$) or $P_{i}x = x$ for exactly
one $i$ (in which case ($\vee_{i} P_{i})x = (\sum_{i} P_{i})x = x$).  Since
every vector in the image space of $W$ is a linear combination of vectors in
the subspaces corresponding to the $X \in {\bf X}$, the claim implies
$(\vee_{i}P_{i})W = (\sum_{i}P_{i}) W$.  This yields countable additivity:
\begin{eqnarray*}
\mu (\cup_{i} S_{P_{i}}) & = & \mu (S_{\vee_{i}P_{i}})  \; \equiv \; {\rm Prob}_{W}(\vee_{i} P_{i} = 1) \; = \; {\rm Tr} ((\vee_{i}P_{i})W) \\
& = & {\rm Tr} ((\sum_{i} P_{i})W) \;
= \; \sum_{i} {\rm Tr} (P_{i} W) 
\; = \; \sum_{i} {\rm Prob}_{W}(P_{i}=1) \\
& \equiv & \sum_{i} \mu (S_{P_{i}}) .
\end{eqnarray*}

Finally (!), we prove that $\mu$ satisfies (1).

Consider any mutually commuting subset $\{ A, B, C , \ldots \}$ of $\overline{{\bf d}}$
with corresponding sets of eigenvalues $\{ \alpha , \beta , \gamma, \ldots \}$.
We have:
\begin{eqnarray*}
{\rm Prob}_{W}(A \in \alpha, B \in \beta, C \in \gamma, \ldots ) & = & {\rm Tr}
(P_{\alpha} P_{\beta} P_{\gamma} \ldots W ) \\
& = &  {\rm Prob}_{W}(P_{\alpha} P_{\beta} P_{\gamma} \ldots = 1) .
\end{eqnarray*}
But since $P_{\alpha}$, $P_{\beta}$, $P_{\gamma}$, $\ldots$ commute and $\overline{{\bf d}}$ is
$^{\ast}$-closed, $P_{\alpha} P_{\beta} P_{\gamma} \ldots$ is a projection in $\overline{{\bf d}}$.
So by definition ${\rm Prob}_{W}(P_{\alpha} P_{\beta} P_{\gamma} \ldots = 1)$
is equal to $\mu \{ \langle .  \rangle \in {\cal F}: \langle P_{\alpha} P_{\beta}
P_{\gamma} \ldots \rangle  = 1 \}$, and we have
\[
{\rm Prob}_{W}(A \in \alpha, B \in \beta, C \in \gamma, \ldots ) =
\mu \{ \langle .\rangle \in {\cal F}: \langle P_{\alpha} P_{\beta} P_{\gamma}
\ldots  =
1\},
\]
which, using the fact that $\langle .\rangle$  is a functional valuation,
gives:
\begin{equation}
\nonumber
\begin{aligned}
{\rm Prob}_{W}(A \in \alpha, B & \in \beta, C \in \gamma, \ldots ) \\
& = \mu \{
\langle .\rangle \in {\cal F}: \langle P_{\alpha} \rangle \langle P_{\beta}
\rangle \langle P_{\gamma} \rangle \ldots = 1 \} \\
& =  \mu \{
\langle .\rangle \in {\cal F}: \langle P_{\alpha} \rangle = \langle P_{\beta}
\rangle = \langle P_{\gamma} \rangle = \ldots = 1 \} \\
& =  \mu \{
\langle .\rangle \in {\cal F}: \langle A \rangle \in \alpha , \langle B \rangle
\in \beta , \langle C \rangle \in \gamma , \ldots \} ,
\end{aligned}
\end{equation}
and (1) is proved.  \ \emph{QED}. \\[1\baselineskip] \indent
It is probably worthwhile to summarize as plainly and briefly as possible
what has happened.  Together, Thms.~2 and 5 show that a set of projection
operators ${\bf d}$ will serve as the basis of a set of (discrete)
definite-valued observables that is $^{\ast}$-\emph{closed} and admits enough
\emph{functional valuations} to represent the quantum statistics for the
observables in the set, if and only if two (logically independent) conditions
are satisfied: \begin{enumerate}
\item ${\bf d} = \underline{{\bf P}^{\prime}}$ for some set of projections
${\bf P}$, and
\item ${\bf d}$ is an $I$-quasiBoolean algebra, \mbox{where $I = \{ P \in {\bf d}\! :
{\rm Prob}_{W}(P=1) =0\}$.}
\end{enumerate}
\renewcommand{\labelenumi}{\theenumi.}
Furthermore, the Corollaries to Thms.~2 and 4 offer a direct method for
constructing such a projection set:  simply specify a set ${\bf X}$ of
mutually orthogonal projections that span a subspace of ${\sf H}$ containing
the image space of $W$.  The resulting ${\bf X}$-form lattice will then
satisfy both (1) and (2).  And notice once more that all the concrete
proposals for sets of definite-valued projections considered in Section 2.3
(save the naive realist's!) are constructed in exactly this way.  In view of
this, it would be nice to know whether \emph{all} sets of projections
satisfying (1) and (2) arise as ${\bf X}$-form lattices with the span of the
mutually orthogonal projections in ${\bf X}$ containing $W$'s image space.

\section{Von Neumann's No-Hidden-Variables Theorem}

At last, we arrive at the infamous theorem.  The essential assumption of the
theorem (the one without which the theorem would not follow, and from which
the theorem \emph{does} follow quite apart from von Neumann's other
assumptions) is stated by Bell [1966] to be: 
\begin{quote} 
\emph{Any real linear combination of any two self-adjoint operators represents
an observable, and the same linear combination of expectation values is the
expectation value of the combination.}
\end{quote}
Let us call this \emph{von Neumann's Principle}.

If the expectation values in question are those prescribed by the quantum
state $W$ of the system, then this principle is unobjectionable.  But von
Neumann's proof came under fire by Bell (and most others following him)
because von Neumann also required his principle to hold for the
dispersion-free states postulated by hidden-variable theories.

A dispersion-free state is one in which there is no statistical spread in the
values of observables, and hence the expectation value of any (discrete)
observable in a dispersion-free state must equal one of its eigenvalues.
Given that, it is trivial to show that von Neumann's principle must fail.
Consider a spin-$1/2$ particle and the linear combination of spin 
observables
$(\sigma_{x} + \sigma_{y}) / \sqrt{2}$ which is, itself, 
the operator
which corresponds to the particle's spin component along the direction
bisecting the $x$ and $y$ directions.  (This example is due to Jammer [1974,
274].)  If the expectations of dispersion-free states are to satisfy von
Neumann's principle, then we must have $\pm 1 = (\pm 1 + \pm
1) / \sqrt{2}$ which is absurd.

Bell's own reason for finding the application of von Neumann's principle
to dispersion-free states implausible places the blame on the incompatibility
of $\sigma_{x}$ and $\sigma_{y}$.  Their incompatibility implies that
$\sigma_{x}$, $\sigma_{y}$, and $(\sigma_{x} + \sigma_{y}) / 
\sqrt{2}$ all
require differently oriented Stern-Gerlach apparati to be measured, and so
there is no logical reason to require the values of these three spin
components, only one of which can be measured at any one time (while the
others' values have to be inferred counterfactually) to conform to von
Neumann's principle.  The only constraint is that for empirical adequacy of a
hidden-variable theory, that principle -- which is a ``quite peculiar
property of quantum states'' (Bell [1966, 449]) -- needs to be reproduced on
averaging over its dispersion-free states.  To make his point, Bell
constructs a simple hidden-variable theory with just that property but with
dispersion-free states that do not satisfy von Neumann's principle.

In our terminology, what Bell questioned was von Neumann's assumption of
faithfulness for the definite values of incompatible observables.  But now we
see that there is another way around the `no-go' theorem -- one which does
not focus exclusive attention on issues of compatibility, and which avoids
making von Neumann's principle out to be merely a peculiarity of quantum
states.  To circumvent the theorem in this sense, we simply drop von
Neumann's \emph{tacit} assumption that every observable must receive a
definite value.  Then Thms.~2 and 5 show that one can actually 
\emph{strengthen} the functional requirements on ${\bf D}$ and its valuations,
and there will still be enough valuations to recover quantum statistics.

Why then did the above spin example go wrong?  Well, if we require that the
set of definite-valued projections ${\bf d}$ include the spectral projections
of $\sigma_{x}$ and $\sigma_{y}$, and if we require that its extension be
$^{\ast}$-closed, then ${\bf d}$ will be a subortholattice of projections in
$L({\sf H}_{2})$ containing one-dimensional projections $P$ and $Q$ that are
neither parallel nor orthogonal (because of the incompatibility of
$\sigma_{x}$ and $\sigma_{y}$).  But such a subortholattice cannot possibly be
an $I$-quasiBoolean algebra, \emph{for any} $I$, since (using Thm.~4's
characterization of such algebras) there is no non-zero projection in $L({\sf
H}_{2})$ contained in or orthogonal to each of $P$ and $Q$.

The conclusion is \emph{not} that von Neumann's principle (or a stronger
principle based on $^{\ast}$-closure rather than just polynomial
$^{\ast}$-closure) must fail for all `hidden-variable' theories.  Rather, all
one can conclude is that the choice of ${\bf d}$ that led to the difficulty
above must be rejected.  And, although in this simple example involving
$\sigma_{x}$ and $\sigma_{y}$ their incompatibility again plays a direct role
in defeating the satisfaction of von Neumann's principle, this is only an
artifact of the two-dimensional case ${\sf H}_{2}$.  Projection sets of ${\bf
X}$-form generate definite-valued sets of observables that satisfy von
Neumann's principle (even with respect to $^{\ast}$-closure), yet in
dimensions higher than 2 they can contain plenty of incompatible projections.
By our characterization theorem for quasiBoolean algebras (Thm.~4), the issue
is not compatibility \emph{per se}, but rather a somewhat `finer' notion:
whether the projections in ${\bf d}$ have sufficiently many \emph{common
eigenvectors} -- the vectors in the ranges of the projections in ${\bf X}$.

Once again, what we learn from von Neumann's theorem is not that
`hidden-variable' theories must give up functional valuations for
non-com\-muting observables, but that they must be more discriminating in what
observables they count as `definite-valued' (i.e.\ having dispersion-free
values).  And since \emph{both} requirements (1) and (2) are satisfied by
${\bf X}$-form projection sets, examples of which include a number of modal
interpretations and the orthodox interpretation, these interpretations
provide a clear existence proof that `hidden-variable' theories \emph{can
indeed} be more discriminating while conforming to von Neumann's principle --
they do not have to simply adopt a naive realism whereby \emph{every}
observable has a definite-value.

The same conclusion spells the demise of the no-go theorems of Jauch and
Piron [1963] and of Kochen and Specker [1967].  In particular, the latter's
theorem weakens von Neumann's principle so that it only carries commitment to
the idea that the set of definite-valued projections is \emph{compatible}
polynomial $^{\ast}$-closed and its dispersion-free states prescribe faithful
valuations respecting only the polynomial functional relations between 
\emph{compatible} observables.  But since we have shown that there is plenty
of room for an interpretation to endorse even a \emph{stronger} version of
von Neumann's original principle based on $^{\ast}$-closure, and we have seen
that eschewing incompatibility is not the ultimate reason for an
interpretation's success in that endeavour, strengthening von Neumann's
(alleged) no-go theorem so that it is sensitive to issues about compatibility
loses its point.

\section*{Acknowledgments}

We would like to thank Pieter~Vermaas for his extremely perceptive
editorial comments on the penultimate draft of this paper.

Jason~Zimba would like to thank the Institute of the Foundations 
of the Natural
Sciences of Utrecht University for its support in enabling him to 
attend this conference. He would also like to thank 
Professor~Martin~Jones for providing many perceptive criticisms, 
and for offering his steady encouragement during the writing of 
this article. Finally, he would like to thank \mbox{Mr.\ Andrew}
Charman 
for a number of insightful comments and fruitful discussions.

Rob~Clifton would like to thank the Social Sciences and Humanities 
Research Council of Canada for funding to attend this conference.

\end{document}